\documentstyle[12pt,epsf]{article}

\newlength{\dinwidth}
\newlength{\dinmargin}
\newcommand{\resection}[1]{\setcounter{equation}{0}\section{#1}}

\begin{document}

\begin{center}
  \begin{Large}
  \begin{bf}
UNIVERSIT\'E DE GEN\`EVE \\
  \end{bf}
  \end{Large}
\smallskip
  \begin{small}
SCHOLA GENEVENSIS MDLIX
  \end{small}
\end{center}
\goodbreak
\begin{center}
\bigskip\vbox
{
\vskip 3truecm
\noindent
\includegraphics{unige.ps}
\vskip 2truecm
\noindent
}
\end{center}
\bigbreak
\begin{center}
  \begin{large}
  \begin{bf}
 DEGENERATE BESS MODEL: the possibility\\
of a low energy strong electroweak sector$^*$\\
  \end{bf}
  \end{large}
  \vspace{1.2cm}
  \begin{large}
R. Casalbuoni$^{(a,b)}$, A. Deandrea$^{(c)}$, S. De Curtis$^{(b)}$\\
D. Dominici$^{(a,b)}$, R. Gatto$^{(c)}$,
M. Grazzini$^{(d,e)}$\\
  \end{large}
\vspace{0.7cm}
$(a)$ Dipart. di Fisica, Univ. di Firenze, Largo E. Fermi, 2; I-50125
Firenze\\
$(b)$ I.N.F.N., Sezione di Firenze, Largo E. Fermi, 2; I-50125
Firenze\\
$(c)$ D\'ept. Phys. Th\'eorique, Univ. de Gen\`eve, 24 quai E.-Ansermet,
CH-1211 Gen\`eve 4\\
$(d)$ Dipart. di Fisica, Univ. di Parma, Viale delle Scienze; I-43100 Parma\\
$(e)$ I.N.F.N., Gruppo Collegato Parma, Viale delle Scienze; I-43100 Parma\\
\end{center}
\vspace{1.5cm}
\begin{center}
UGVA-DPT 1995/10-906\\
hep-ph/9510431\\
October 1995
\end{center}
\vspace{3cm}
\noindent
$^*$ Partially supported by the Swiss National Foundation\\
\newpage
\thispagestyle{empty}
\begin{quotation}
\vspace*{5cm}
\begin{center}
\begin{Large}
\begin{bf}
ABSTRACT
\end{bf}
\end{Large}
\end{center}
\vspace{1cm}
\noindent
We discuss possible symmetries of effective theories describing spinless and
spin 1 bosons, mainly to concentrate on an intriguing phenomenological
possibility: that of a hardly noticeable strong electroweak sector at
relatively low energies. Specifically, a model with both vector and axial
vector strong interacting bosons may possess a discrete symmetry imposing
degeneracy of the two sets of bosons (degenerate BESS model).
In such a case its effects at low energies become almost invisible and the
model easily passes all low energy precision tests.
The reason lies essentially in the fact that the model automatically
satisfies decoupling, contrary to models with only vectors.
For large mass of the degenerate spin one bosons the model becomes
identical at the classical level to the standard model taken in the limit of
infinite Higgs mass.
For these reasons we have thought it worthwhile to fully develop the model,
together with its possible generalizations,
and to study the expected phenomenology. For instance, just because
of its invisibility at low energy, it is conceivable
that degenerate BESS has low mass spin one states and gives quite visible
signals at existing or forthcoming accelerators.
\end{quotation}
\newpage

\newcommand{\f}[2]{\frac{#1}{#2}}
\newcommand{\be}{\begin{equation}}
\newcommand{\ee}{\end{equation}}
\newcommand{\bea}{\begin{eqnarray}}
\newcommand{\eea}{\end{eqnarray}}
\newcommand{\nn}{\nonumber}
\newcommand{\dd}{\displaystyle}
\newcommand{\cdt}{c_{2\theta}}
\newcommand{\sdt}{s_{2\theta}}
\def\vmu{{\bf V}_\mu}
\def\amu{{\bf A}_\mu}
\def\lmu{{\bf L}_\mu}
\def\rmu{{\bf R}_\mu}
\def\rmus{{\bf R}^\mu}
\def\gs{{g''}}
\def\dmu{\partial_\mu}
\def\dnu{\partial_\nu}
\def\dmus{\partial^\mu}
\def\dnus{\partial^\nu}
\def\gp{g'}
\def\gpt{{{\tilde g}^\prime}}
\def\gptd{\tilde g^{\prime 2}}
\def\ggs{\frac{g}{\gs}}
\def\eps{{\epsilon}}
\def\tr{{\rm {tr}}}
\def\V{{\bf{V}}}
\def\W{{\bf{W}}}
\def\Y{{\bf{Y}}}
\def\Yt{\tilde{\bf {Y}}}
\def\tW{\tilde W}
\def\tY{\tilde Y}
\def\tL{\tilde L}
\def\tR{\tilde R}
\def\L{{\cal L}}
\def\s{s_\theta}
\def\c{c_\theta}
\def\gt{\tilde g}
\def\et{\tilde e}
\def\At{\tilde A}
\def\Zt{\tilde Z}
\def\Wt{\tilde{\bf {W}}}
\def\Lt{{\bf {L}}}
\def\Rt{{\bf {R}}}
\def\st{\tilde s_\theta}
\def\ct{\tilde c_\theta}
\def\A{\bf A}
\def\Z{\bf Z}
\def\Wpt{\tilde W^+}
\def\Wmt{\tilde W^-}
\setcounter{page}{1}

\resection{Introduction}

In a first part of this work we shall give a general discussion of possible
properties of low-energy effective theories which describe light pseudoscalar
mesons, vector and axial vector mesons, as for instance the bosonic sector of
low-energy QCD. Indeed QCD itself may be a testing ground for one particular
specification among the low energy effective theories to be discussed.
However
our main interest here will not be QCD, but rather an alternative possible
specification of the low energy theory which may be relevant for an effective
description of the phenomenology arising from schemes of strong
electroweak breaking.

The bulk of this work will be devoted to the formulation of such a highly
symmetric form of low energy effective theory and to the derivation of the
very remarkable electroweak phenomenology that it would originate. In a
simple model one would think of Goldstone bosons absorbed to give masses to
$W$ and $Z$ and, besides, vector and axial vector resonances as the most
visible manifestations at low energy of the strong interacting sector.

We shall call $G$ the symmetry group of the theory, spontaneously broken, of
which the pseudoscalars are the Goldstone bosons. The vector and axial vector
mesons will transform under the unbroken subgroup $H$ of $G$. In the sense of
the method used by CCWZ \cite{ccwz} the vector and axial vector mesons can be
treated as matter fields.

It will be a formal expedient to consider the new vector and axial vector
fields as gauge bosons of a local symmetry $H'$, which is spontaneously
broken. The local symmetry $H'$ is usually referred to as hidden symmetry
\cite{bando}, \cite{bala}.
The spin-one bosons acquire their mass, in this description, by absorption of
the would-be Goldstones related to the spontaneous breaking of $H'$. Indeed
the peculiar feature of this approach is the explicit presence of these
modes. The symmetry group gets enlarged to $G'=G \otimes H'$, where $G$ is
global and $H'$ local. The diagonal subgroup of $H\otimes H'$ ($H' \supseteq
H$), formally isomorphic to $H$, is called $H_D$ and it is the invariance
group of the vacuum.

We shall mainly consider the case $G=SU(2)_L \otimes SU(2)_R$,
$H'=SU(2)_L \otimes SU(2)_R$ and $H_D=SU(2)_V$, where $H_D$ is the diagonal
$SU(2)$ subgroup of $G'$. The group $G'$ breaks down spontaneously to $H_D$
and
gives rise to nine Goldstones. Of these, six are absorbed by the vector and
axial vector bosons, which are triplets of $SU(2)_V$. The three Goldstones
remaining in the spectrum are massless, at least as long as a part of
$G$ is not promoted
to a local group. This situation is discussed in \cite{bando} for QCD and
in \cite{assiali} in the context of dynamical electroweak
symmetry breaking.

The detailed study of the symmetries of the effective theory shows however
that in special cases the resulting symmetry can be larger than the one
requested by the construction. For particular choices of the parameters, a
maximal symmetry $[SU(2)\otimes SU(2)]^3$ can be realized for
the low energy effective lagrangian of the pseudoscalar,
vector-, and axial vector- bosons. Two choices are
possible. One can be seen as the natural generalization of the vector
symmetry of Georgi \cite{georgi} for the case when axial vector mesons are
also included in addition to the vector mesons of the vector-symmetry; this
choice has been considered in ref. \cite{vecax}.

The second choice is the one on which we shall focus here. It may be useful
in relation to schemes of strong electroweak breaking. In fact it has the
interesting feature of allowing for a low energy strong electroweak resonant
sector while satisfying at the same time the severe constraints from low
energy experiments, particularly from LEP/SLC. As such it offers
possibilities of experimental test even with future or existing
machines of relatively low energy. The phenomenological implications will be
a substantial part of our discussion below.

The type of realization of the maximal symmetry $[SU(2)\otimes SU(2)]^3$ on
which we shall focus in this work automatically implies degenerate vector and
axial vector mesons which have the property of not coupling to the
pseudoscalars. The model, after introducing the gauge couplings of the
electroweak vector bosons, will be
called degenerate BESS (BESS stands for Breaking Electroweak
Symmetry Strongly). We shall study in detail its phenomenology. We stress
immediately its main property and what makes it so attractive: in degenerate
BESS, also when extended to a larger initial symmetry (for instance $SU(8)$
in place of $SU(2)$), one generally derives that all deviations in the low
energy parameters from their standard model (SM) values are strongly
suppressed. This would make it possible that a strong electroweak sector at
relatively low energies exists within the precision of electroweak tests,
such that it may be accessible with existing accelerators (Tevatron)
or with accelerators in construction or
projected for the near future. In fact one can show that the lagrangian of
degenerate BESS becomes identical to that of the standard model
(except for the Higgs sector) for
sufficiently large mass of the degenerate vector and axial vector mesons. In
other words, different form ordinary BESS \cite{bess}, where
such a high mass decoupling is not satisfied, the decoupling occurs in
degenerate BESS.

The decoupling theorem valid for degenerate BESS requires an
accurate study of the contributions of momentum dependent terms to virtual
effects of the heavy particles. One can then evaluate such virtual effects
for
LEP and Tevatron, and subsequently examine what modification of the trilinear
gauge couplings may be visible at higher energy $e^+e^-$ colliders. The
discussion requires careful redefinition of the physical constants in terms
of
the parameters of the effective lagrangian. As well known, in the low energy
limits one can parameterize the modifications due to the heavy sector in
terms
of three independent parameters ($\Delta r_W$, $\Delta k$, $\Delta \rho$, or
equivalently $\epsilon_1$, $\epsilon_2$, $\epsilon_3$). Radiative corrections
have also to be taken into account. The result of this analysis, that we
shall present first, shows that in degenerate BESS relatively light resonances
are indeed compatible with the electroweak data, as given by LEP and Tevatron.

Besides studying the virtual effects of the heavy resonances we shall also
discuss their direct production. To this aim full
couplings to fermions and the trilinear couplings among the physical bosons
are needed; physical quantities must be carefully identified by renormalizing
the occurring expressions and choosing the electric charge, the $Z$ mass, and
the Fermi constant as physical inputs.
Our phenomenological applications include discussion of the properties of the
heavy resonances (masses, partial widths) and studies of their effects at
Tevatron, at $e^+e^-$ colliders, and at hadron colliders. The Tevatron limits
on $W'$ can be used to limit the parameter space of degenerate BESS. A
feature of degenerate BESS, as compared to BESS with only vector resonances,
comes from the absence of direct coupling of the new resonances to the
longitudinal weak gauge bosons. This implies larger widths into
fermion pairs as compared to widths into pairs of weak gauge bosons.
Comparison
of the limits one can get from CDF to those from LEP shows that CDF is more
efficient in limiting low resonance masses while LEP is more efficient for
larger masses.

The sensitivity of degenerate BESS at LEP2 and higher energy linear colliders
will be discussed by comparing cross-sections and asymmetries in the
fermionic pair channels and $WW$ channel between the model and SM. For LEP2
the general conclusion will be that the bounds on the model would not be much
stronger that those from LEP. Substantial improvements are expected from a
500 $GeV$ $e^+e^-$ collider for 20 $fb^{-1}$, even without beam polarization.
The $WW$ final state
does not contribute in an important way to the attainable bounds which come
essentially from the fermion channels alone (this is a characteristics of
degenerate BESS, as already said). Hadron colliders would be complementary to
$e^+e^-$ colliders and hopefully will allow for direct study of the new
resonant states. For instance, a charged resonance with
 mass of 500 $GeV$ could give at
LHC a spectacular signal. Higher masses up to 1.5 $TeV$ would still give
significant signals. Degenerate BESS would thus be comparatively much more
evident than ordinary BESS, and probably than any other strong electroweak
model not sharing its peculiar symmetry properties.

In section 2 we recall briefly the effective lagrangian formalism we employ
in
describing vector and axial vector resonances. In section 3 we introduce the
lagrangian describing our model with extended symmetry of degenerate vector
and
axial vector resonances. In sections 4, 5, 6 and 7 we consider the low energy
limit of the model, integrating out the new vector and axial vector bosons,
both considering the leading order and the next-to-leading order.
Implication for the LEP observables are derived as well as other virtual
effects of the heavy particles that may be relevant at low energy. In
sections
8, 9 and 10 we consider the possibility of direct production of the heavy
resonances, so the predictions of the complete lagrangian of the theory are
derived, such as mass formulae and eigenstates of the new particles,
fermionic
and trilinear couplings. In section 11 the physical quantities of the model
are
identified with the usual renormalization procedures. In section 12 width
formulae of relevance in the study of the model are derived. In sections 13,
14 and 15 we discuss the phenomenological implications of the model at
present
and future high energy accelerators.

\resection{Extended Vector-Axial Symmetry}

Let us consider the following group structure: $G=SU(2)_L\otimes SU(2)_R$,
$H'=SU(2)_L\otimes SU(2)_R$ and $H_D=SU(2)_V$, as already stated in the
introduction. The nine Goldstone bosons resulting from the spontaneous
breaking
of $G'=G\otimes H'$ to $H_D$, can be described by three independent
$SU(2)$ elements: $L$, $R$ and $M$, transforming with respect to $G$ and
$H'$ as follows
\be
L'= g_L L h_L,~~~~~~R'= g_R R h_R,~~~~~~M'= h_R^\dagger M h_L
\ee
with $g_{L,R}\in G$ and $h_{L,R}\in H'$. Moreover we shall require the
invariance under the discrete left-right transformation, denoted by $P$
\be
P:~~~~~~~L\leftrightarrow R,~~~~~M\leftrightarrow M^\dagger
\ee
which ensures that the low energy theory is parity conserving.

If we ignore the transformations of eq. (2.1), the largest possible global
symmetry of the low energy theory is given by the requirement of maintaining
for the transformed variables $L'$, $R'$ and $M'$, the character of $SU(2)$
elements.

The maximal symmetry is given by the group $G_{max}=[SU(2)\otimes SU(2)]^3$,
consisting of three independent $SU(2)\otimes SU(2)$ factors, acting on each
of
the three variables separately. It happens that, for specific choices of the
parameters of the theory, the symmetry $G'$ gets enlarged to $G_{max}$.

The most general $G'\otimes P$ invariant lagrangian is given by
\cite{assiali}
\be
{\cal L}_G=-\frac{v^2}{4} [a_1 I_1 + a_2 I_2 + a_3 I_3 + a_4 I_4]
\label{lg}
\ee
plus the kinetic terms ${\cal L}_{kin}$.  The terms $I_i$ ($i=1,...4$) are
given by:
\bea
I_1&=&tr[(V_0-V_1-V_2)^2]\nn\\
I_2&=&tr[(V_0+V_2)^2]\nn\\
I_3&=&tr[(V_0-V_2)^2]\nn\\
I_4&=&tr[V_1^2]
\eea
and
\bea
V_0^\mu&=&L^\dagger D^\mu L\nn\\
V_1^\mu&=&M^\dagger D^\mu M\nn\\
V_2^\mu&=&M^\dagger(R^\dagger D^\mu R)M
\eea
The covariant derivatives are
\bea
D_\mu L&=&\partial_\mu L -L \lmu\nn\\
D_\mu R&=&\partial_\mu R -R \rmu\nn\\
D_\mu M&=&\partial_\mu M -M \lmu+\rmu M
\eea
The kinetic term is
\be
{\cal L}_{kin}=\frac{1}{\gs^2} tr[F_{\mu\nu}({\bf L})]^2+
	 \frac{1}{\gs^2}  tr[F_{\mu\nu}({\bf R})]^2
\ee
where $\gs$ is the gauge coupling constant for the gauge fields $\lmu$ and
$\rmu$,
\be
F_{\mu\nu}({\bf L})=\partial_\mu{\bf L}_\nu-\partial_\nu{\bf L}_\mu+
	 [\lmu,{\bf L}_\nu]
\ee
and the same definition holds for $F_{\mu\nu}({\bf R})$.
In eq. (\ref{lg}) $v$ represents a physical scale related to the spontaneous
symmetry breaking of the theory, depending on the particular context under
investigation.

The quantities $V_i^\mu~~(i=0,1,2)$ are invariant under the global symmetry
$G$
and covariant under the gauge group $H'$
\be
(V_i^\mu)'=h_L^\dagger V_i^\mu h_L
\ee
Using the $V_i^\mu$ one can build six independent quadratic invariants,
which reduce to the four $I_i$ listed above, when parity conservation is
required.

For generic values of the parameters $a_1,~a_2,~a_3,~a_4$, the lagrangian
${\cal L}$ is invariant under $G'\otimes P=G\otimes H'\otimes P$.
There are however special choices which enhance the symmetry group
\cite{vecax}.

The case of interest for the electroweak sector is provided by the choice:
$a_4=0$, $a_2=a_3$. In order to discuss the symmetry properties it is useful
to
observe that the invariant $I_1$ can be re-written as follows
\be
I_1=-tr(\partial_\mu U^\dagger \partial^\mu U)
\ee
with
\be
U=L M^\dagger R^\dagger
\ee
and the lagrangian becomes
\be
{\cal L}_G=\frac{v^2}{4}\{a_1~ tr(\partial_\mu U^\dagger \partial^\mu U) +
			 2~a_2~ [tr(D_\mu L^\dagger D^\mu L)+
			  tr(D_\mu R^\dagger D^\mu R)]\}
\label{lg1}
\ee
Each of the three terms in the above expression
is invariant under an independent $SU(2)\otimes SU(2)$
group
\be
U'=\omega_L U \omega_R^\dagger,~~~~~~L'= g_L L h_L,~~~~~~R'= g_R R h_R
\ee
Moreover, whereas these transformations act globally on the $U$
fields, for the variables $L$ and $R$, an $SU(2)$ subgroup is
local. The overall symmetry is $G_{max}=[SU(2)\otimes SU(2)]^3$, with a part
$H'$ realized as a gauge symmetry.

The field redefinition from the variables $L$, $R$ and $M$ to
$L$, $R$ and $U$ has no effect on the physical content of the theory.

The extra symmetry related to the independent transformation
over the $U$ field, can also be expressed in terms of the original
variable $M$. Indeed the lagrangian of eq. (\ref{lg}), for $a_4=0,~a_2=a_3$,
possesses the additional invariance
\be
L'=L,~~~~~~R'=R,~~~~~~M'=\Omega_R M \Omega_L^\dagger
\ee
with
\be
\Omega_L=L^\dagger \omega_L L,~~~~~~\Omega_R=R^\dagger \omega_R R
\ee

By expanding the lagrangian in eq. (\ref{lg}) in powers of the Goldstone
bosons
one finds, as the lowest order contribution, the mass terms for the vector
and axial vector mesons
\be
{\cal L}_G = -\frac{v^2}{4}[a_2~ tr(\lmu+\rmu)^2 + a_2~
	 tr(\lmu-\rmu)^2]+\cdots
\label{lge}
\ee
where the dots stand for terms at least linear in the Goldstone modes.
The mixing between $\lmu$ and $\rmu$ is vanishing, and the states are
degenerate in mass. Therefore, in the following we will not work with
vector and axial vector combinations but with the $\lmu$ and $\rmu$
components.
Moreover, as it follows from eq. (\ref{lg1}), the longitudinal modes
of the $\lmu$ and $\rmu$  fields are entirely
provided by the would-be Goldstone bosons in $L$ and $R$. This means
that the pseudoscalar particles remaining as physical states in the
low energy spectrum are those associated to $U$. They in turn can
provide the longitudinal components to the $W$ and $Z$ particles,
in an effective description of the electroweak breaking sector.

\resection{The degenerate BESS model}

We now consider the coupling of the model to the electroweak
$SU(2)_L\otimes U(1)_Y$ gauge fields via the minimal substitution
\bea
D_\mu L &\to& D_\mu L+ {\Wt}_\mu L\nn\\
D_\mu R &\to& D_\mu R+ {\Yt}_\mu R\nn\\
D_\mu M &\to& D_\mu M
\eea
where
\bea
\Wt_\mu&=&i\gt {\tilde W}_\mu ^a\f{\tau^a}{2}\nn\\
\Yt_\mu&=&i\gpt
 {\tilde Y}_\mu\f{\tau^3}{2}\nn\\
\Lt_\mu&=&i\f{\gs}{\sqrt{2}} L_\mu ^a\f{\tau^a}{2}\nn\\
\Rt_\mu&=&i\f{\gs}{\sqrt{2}} R_\mu ^a\f{\tau^a}{2}
\eea
with $\gt$, $\gpt$ the $SU(2)_L\otimes U(1)_Y$
gauge coupling constant and $\tau^a$ the Pauli matrices.

By introducing the canonical kinetic terms for $W_\mu^a$ and $Y_\mu$ we get

\be
\L=-\f{v^2}{4}\Big[ a_1 \tr(\Wt_\mu-\Yt_\mu)^2
+2 a_2 \tr(\Wt_\mu-{\Lt}_\mu)^2
+2 a_2 \tr(\Yt_\mu-\Rt_\mu)^2\Big]
+\L^{kin}(\Wt,\Yt,\Lt,\Rt)
\ee
\bea
\L^{kin}(\Wt,\Yt,\Lt,\Rt)&=&
\f{1}{2 \gt^2}\tr[F^{\mu\nu}(\Wt)F_{\mu\nu}(\Wt)]
   +\f{1}{2 {\tilde g}^{\prime 2}}\tr[F^{\mu\nu}(\Yt)
F_{\mu\nu}(\Yt)]\nn\\
&+&\f{1}{ \gs^2}\tr[F^{\mu\nu}(\Lt)F_{\mu\nu}(\Lt)]+
\f{1}{ \gs^2}\tr[F^{\mu\nu}(\Rt)F_{\mu\nu}(\Rt)]
\eea

We have used tilded quantities to remember that, due to the effects
of the $\Lt$ and $\Rt$ particles, they are not the physical parameters
and fields.
In the next sections we will derive the relations between the tilded
quantities and the physical ones.

 It is natural to think about
the model we are considering as a perturbation around the SM picture.
The SM relations are obtained in the limit $\gs \gg {\tilde g},
{\tilde g}'$. Actually,
for a very large $\gs$, the kinetic terms for the fields $\lmu$ and $\rmu$
drop out, and ${\cal L}$ reduces to the first term in eq. (3.3).
This term reproduces precisely the mass term
for the ordinary gauge vector bosons in the SM, provided we
identify the combination $v^2 a_1$ with $1/(\sqrt{2} G_F)$, $G_F$
being the Fermi constant.
Therefore in the following we will assume \cite{assiali}
\be
a_1=1
\ee
Finally let us consider the fermions of the SM and denote them by $\psi_L$
and $\psi_R$. They couple to $\Lt$ and $\Rt$ via the mixing with the
standard $\Wt$ and $\Yt$:
\bea
\L_{fermion} &=& \overline{\psi}_L i \gamma^\mu\Big(\dmu+\Wt_\mu+
		      \f{i}{2}\gpt(B-L){\tilde Y}_\mu\Big){\psi}_L\nn\\
     &+&\overline{\psi}_R i \gamma^\mu\Big(\dmu+\Yt_\mu+\f{i}{2}\gpt
		      (B-L){\tilde Y}_\mu\Big){ \psi}_R
\eea
where $B(L)$ is the baryon (lepton) number, and
\be
\psi =\left(
\begin{array}{c}
\psi_u\\
\psi_d\end{array}
\right)
\ee
In addition, we also expect direct couplings to the new vector bosons
since they are allowed by the symmetries of $\L$ \cite{bess},
but, for simplicity, this possibility will not be considered here.

\resection{The low energy limit}

We want to study the effects of the $\Lt$ and $\Rt$
particles in the low energy limit \cite{anichini}.
This can be done by eliminating the $\Lt$ and $\Rt$ fields with the solution
of their equations of motion for $M_{L,R}\to\infty$. In fact in this
limit the kinetic terms of the new resonances are negligible.
Neglecting electromagnetic corrections the common mass of the resonances
is  given by $M^2\simeq a_2 (v^2/4) \gs^2$.

The $M\to\infty$ limit
can be taken in two different ways (we consider $v$ fixed to its experimental
value): by sending $\gs\to\infty$ with $a_2$ fixed or by fixing $\gs$
and sending $a_2\to\infty$. In the first case  the $\Lt$
and $\Rt$ bosons trivially decouple.
We will now show that also in the second case we have decoupling due
to the extended symmetry $[SU(2)\otimes SU(2)]^3$.

Let us solve the equations of motion for $\Lt$ and $\Rt$ in this limit.
We get
\bea
\Lt_\mu=\Wt_\mu\nn\\
\Rt_\mu=\Yt_\mu,
\eea
where the last equation means that only the third isospin component of
$\Rt$ is different from zero.
By substituting these equations in the total lagrangian (see eqs. (3.3)
and (3.6) for the case of $SU(2)$) we get
\bea
\L_{eff} &=& -\f{v^2}{4}\tr(\Wt_\mu-\Yt_\mu)^2\nn\\
&+& \f{1}{2 \gt^2}\tr[F^{\mu\nu}(\Wt)F_{\mu\nu}(\Wt)]
   +\f{1}{2 {\tilde g}^{\prime 2}}\tr[F^{\mu\nu}(\Yt)
F_{\mu\nu}(\Yt)]\nn\\
&+&\f{1}{ \gs^2}\tr[F^{\mu\nu}(\Wt)F_{\mu\nu}(\Wt)]+
\f{1}{ \gs^2}\tr[F^{\mu\nu}(\Yt)F_{\mu\nu}(\Yt)]\nn\\
    &+& \L_{fermion}
\eea

 From eq. (4.2) we see that the effective contribution of the $\Lt$
and $\Rt$ particles
give additional terms to the kinetic terms of the standard $\Wt$ and $\Yt$.
By the following redefinition of the coupling constants
\bea
\f 1 {2 g^2} &=& \f 1 {2 \gt^2} +\f 1 {\gs^2}\nn\\
\f 1 {2 g^{\prime 2}} &=& \f 1 {2 {\tilde g}^{\prime 2}} +\f 1 {\gs^2}
\eea
the effective lagrangian becomes identical to the one of the SM
(except for the Higgs sector) showing
the decoupling of the theory in the limit $M\to\infty$.
Let us comment about this fact. If one starts from the most general
lagrangian, eq. (2.3), gauged according to eq. (3.1), in the limit
$M\to\infty$ the solutions of the equations of motion for $\Lt$ and
$\Rt$ are the following
\bea
\Lt_\mu &=& \f 1 2 (1+z)\Wt_\mu+ \f 1 2 (1-z)\Yt_\mu\nn\\
\Rt_\mu &=& \f 1 2 (1-z)\Wt_\mu+ \f 1 2 (1+z)\Yt_\mu
\eea
with
\be
z=\f {a_3}{a_3+a_4}
\ee
By substituting in the lagrangian
\bea
\L_{eff} &=& -\f{v^2}{4}\tr(\Wt_\mu-\Yt_\mu)^2\nn\\
&+& \left(\f{1}{2 \gt^2}+\f 1 {2\gs^2}(1+z^2)\right)
\tr[F^{\mu\nu}(\Wt)F_{\mu\nu}(\Wt)]\nn\\
  & +&\left(\f{1}{2 {\tilde g}^{\prime 2}}+\f 1 {2\gs^2}(1+z^2)\right)
\tr[F^{\mu\nu}(\Yt)
F_{\mu\nu}(\Yt)]\nn\\
&+&
\f{1}{ \gs^2}(1-z^2)\tr[F^{\mu\nu}(\Yt)F_{\mu\nu}(\Wt)]\nn\\
    &+& \L_{fermion}
\eea
We see that all the corrections to the SM lagrangian depend on the value of
the parameter $z$. In the case we are considering, $a_4=0$, $a_2=a_3$,
we have $z=1$, so the corrections
vanish. Notice that the requirement $z=1$ implies only $a_4=0$, so the
corrections would be zero also for $a_2\not=a_3$, but in this case we
would
not have an enlargement of the symmetry and correspondingly there would
be no protection from radiative corrections.
The corrections would vanish also for $z=-1$ but again
this case does not correspond to an extra-symmetry.

We note that the
decoupling remains also valid in the general case of
an extended symmetry $[SU(N)\otimes SU(N)]^3$ provided of a suitable
redefinition of the $SU(2)_L\otimes U(1)_Y$ gauge coupling constants.
When $SU(N)\otimes SU(N) \supset SU(3)$ and one considers also the
$SU(3)_{color}$ gauging, a redefinition of the strong gauge coupling constant
$g_s$
is necessary as well. This happens for instance in the model
considered in ref. \cite{bessu8}.
In the case of an extended
symmetry $[SU(8)\otimes SU(8)]^3$   we find
\bea
\f 1 {2 g^2} &=& \f 1 {2 \gt^2} +\f 1 {\gs^2}\nn\\
\f 1 {2 g^{\prime 2}} &=& \f 1 {2 {\tilde g}^{\prime 2}} +\f 5 3 \f 1
{\gs^2}\nn\\
\f 1 {4 g_s^2} &=& \f 1 {4 {\tilde g}_s^2} +\f 1 {2\gs^2}
\eea

\resection{The low energy limit, next-to-leading order}

Since the degenerate BESS model is indistinguishable
from the SM at the leading order in the low energy limit $(M\to\infty)$
let us consider the solution of the classical equations of motion for the
$\Lt$ and $\Rt$ fields by
retaining also terms of the order $q^2/M^2$.
As in sect. 4  we will eliminate the $\Lt$ and $\Rt$ fields
with the solutions
of their equations of motion and we will consider the virtual effects
of the heavy particles.
We will study the
effective theory by considering the limit $\gs\to\infty$
with corrections up to order $(1/\gs)^2$.

Let us solve the equations of motion for $\Lt$ and $\Rt$ in this limit.
We get
\bea
\Lt_\nu=\left(1-\f{\Box}{M^2}\right)\Wt_\nu+\Delta \Lt_\nu \nn\\
\Rt_\nu=\left(1-\f{\Box}{M^2}\right)\Yt_\nu+\Delta\Rt_\nu
\eea
with
\bea
\Delta\Lt^\nu &=& \f 1 {M^2} \Big( \dnus \dmu  \Wt^\mu-\dmu
[\Wt^\mu, \Wt^\nu]-[\Wt_\mu, F^{\mu\nu}(\Wt)]\Big)\nn\\
\Delta\Rt^\nu &=& \f 1 {M^2} \Big( \dnus \dmu  \Yt^\mu\Big)
\eea
where the  equation for $\Rt$ means that only the third isospin
component of the field  is different from zero.
$\Delta\Lt_\mu$ and $\Delta\Rt_\mu$ contain
linear terms proportional to the divergences of the fields and
$\Delta\Lt_\mu$ contains also
bilinear and trilinear
terms which do not
affect the self-energies and contribute to the anomalous trilinear and
quadrilinear couplings.

We will examine the virtual effects of the $\Lt$ and $\Rt$ fields on the
observables. In particular, in the next section, we will focus on the
physics at LEP and Tevatron, for which the modifications due to the
heavy particles affect the self-energies only. After we will discuss the
modifications in the trilinear gauge couplings which will be studied at
future $e^+e^-$ colliders.

To discuss the LEP physics we neglect $\Delta\Lt_\mu$ and
$\Delta\Rt_\mu$
in the solutions (5.1).
By substituting in the lagrangian (3.3) we get for the
bilinear part (neglecting again divergences of the vector fields)
\bea
\L^{(2)}_{eff} &=&
 -\f{1}{4} (1+z_\gamma)\At_{\mu\nu} \At^{\mu\nu}
-\f{1}{2} (1+z_w)\Wpt_{\mu\nu} {\tilde W}^{\mu\nu -}
-\f{1}{4}(1+z_z) \Zt_{\mu\nu} \Zt^{\mu\nu}\nn\\
&+&\f 1 2 z_{z\gamma}\At_{\mu\nu}\Zt^{\mu\nu}
 + {\tilde M}_W^2  {\tilde W}^{\mu+}{\tilde W}^-_\mu
+\f{1}{2}{\tilde M}_Z^2  \Zt^\mu \Zt_\mu\nn\\
&+&  \f{1}{2M^2} \Big[ \f{z_z}{2} \Zt_{\mu\nu}\Box \Zt^{\mu\nu}
+ \f {z_{\gamma}} 2 \At_{\mu\nu}\Box \At^{\mu\nu}
-z_{z\gamma}\Zt_{\mu\nu}\Box \At^{\mu\nu}
+z_w{\tilde W}^+_{\mu\nu}\Box{\tilde W}^{-\mu\nu}
\Big]
\eea
where
\bea
{\tilde W}_\mu^\pm &=&\f{1}{\sqrt{2}}({\tilde W}_1\mp i {\tilde W}_2)\nn\\
{\tilde W}_\mu^3 &=& \st \At_\mu+\ct \Zt_\mu\nn\\
{\tilde Y}_\mu &=& \ct \At_\mu-\st \Zt_\mu\nn\\
\et &=& \gt\st = \gpt\ct\nn\\
{\tilde M}_W^2&=&\f {v^2} 4 \gt^2\nn\\
{\tilde M}_Z^2&=&\f {{\tilde M}_W^2}{\ct^2}
\eea
 $O_{\mu\nu}=\dmu O_\nu-\dnu O_\mu$, ($O={\tilde W}^\pm,\At,\Zt$), and
\be
z_\gamma = 4 \s^2\Big(\f{g}{\gs}\Big)^2~~~~~
z_w =  2\Big(\f{g}{\gs}\Big)^2~~~~~
z_z =   \f{1+c^2_{2\theta}}{\c^2}\Big(\f{g}{\gs}\Big)^2~~~~~
z_{z\gamma} =  -2 \f{\s}{\c}c_{2\theta}  \Big(\f{g}{\gs}\Big)^2
\ee
Notice that $\gt,\gpt,\et,\st,\ct$ have the same definitions
 as in the SM. As stated before,
 due to the effects of the $\Lt$ and $\Rt$ particles, these
are not the physical quantities in our model.
In eq. (5.5) we have not used the tilded quantities since these parameters
are already of the order of $(1/\gs)^2$.

The corrections to $\L_{SM}$ are $U(1)_{em}$ invariant and produce
a wave-function renormalization of $\At_\mu,\Zt_\mu,{\tilde W}_\mu^\pm$
plus a
mixing term $\At_\mu-\Zt_\mu$. We will absorb these corrections
by a convenient redefinition of the fields. Actually there are only
three renormalization transformations of the fields
$\At_\mu,\Zt_\mu,{\tilde W}_\mu^\pm$ without changing the physics. This
means that three of the four deviations $z_\gamma,z_w,z_z,z_{z\gamma}$ are
non physical.

To identify the physical quantities we define new fields in such a way
to have canonical kinetic terms and to cancel the mixing term
$\At_\mu-\Zt_\mu$. They are the following:
\bea
\At_\mu &=&\Big (1-\f{z_\gamma}{2}(1-  \f\Box {M^2})\Big)
 A_\mu+z_{z_\gamma}\Big(1-\f\Box{M^2}\Big) Z_\mu\nn\\
{\tilde W}^\pm_\mu &=& \Big(1-\f{z_w}{2}(1+\f {M_W^2}{M^2}-
\f {\Box}{M^2})\Big) W_\mu^\pm\nn\\
\Zt_\mu &=& \Big(1-\f{z_z}{2}(1+\f {M_Z^2}{M^2}-
\f {\Box}{M^2})\Big) Z_\mu
\eea
Working at the first order in $1/M^2$ and in $1/\gs^2$, we do not make
distinction in the coefficients of these parameters between "tilded" and
physical quantities.
By substituting in (5.3) we get
\bea
\L^{(2)}_{eff} &=&
 -\f{1}{4} A_{\mu\nu} A^{\mu\nu}
-\f{1}{2} W^+_{\mu\nu} W^{\mu\nu -}
-\f{1}{4}Z_{\mu\nu} Z^{\mu\nu}\nn\\
&+&  {\tilde M}_W^2 \Big(1-z_w(1+\f {M_W^2}{M^2})\Big) W^{\mu+}
 W^-_\mu
+\f{1}{2}{\tilde M}_Z^2\Big(1-z_z(1+\f {M_Z^2}{M^2})\Big)  Z^\mu Z_\mu
\eea
 From which we obtain the values of the physical masses
\bea
M_W^2&=&{\tilde M}_W^2\Big(1-z_w(1+\f {M_W^2}{M^2})\Big)\nn\\
M_Z^2&=&{\tilde M}_Z^2\Big(1-z_z(1+\f {M_Z^2}{M^2})\Big)
\eea
The field renormalization affects also all the couplings of the standard
gauge bosons to the fermions. By separating the charged and
the neutral fermionic sector and substituting eq. (5.6) in $\L_{fermion}$
given in (3.6) we get
\bea
\L_{eff}^{charged} &=& -\f{\et}{\sqrt{2} \st}
     \overline\psi_d
     \gamma^\mu\f{1-\gamma_5}{2}\psi_u
\Big(1-\f{z_w}{2}(1+\f{M_W^2}{M^2}-\f\Box{M^2})\Big)
W^-_\mu +~h.c.\\
\L_{eff}^{neutral} &=& -\f{\et}{\st \ct}
      \overline\psi
     \gamma^\mu\Big[ T^3_L \f{1-\gamma_5}{2}\nn\\
&-&Q \st^2 \Big(1
      -\f{\ct}{\st}z_{z \gamma}(1-\f{\Box_Z}{M^2})\Big)\Big]
       \psi~
\Big(1-\f{z_z}{2}(1+\f{M_Z^2}{M^2}-\f\Box{M^2})\Big)
Z_\mu\nn\\
     & &  - \et
	\overline\psi \gamma^\mu Q \psi~
\Big(1-\f{z_\gamma}{2}(1-\f\Box{M^2})\Big)A_\mu
\eea
with the standard definitions:
\bea
Q &=& \f{\tau^3}{2} +\f{B-L}{2}\nn\\
T^3_L \psi_L &=& \f{\tau^3}{2}\psi_L~~~~~~T^3_L \psi_R =0
\eea
and $\Box_Z$ operates only on the $Z$ field.

The physical constants as the electric charge,
the Fermi constant
and the mass of the $Z$, which are the input parameters for
 the physics at LEP, must be redefined in terms of the parameters appearing
in our effective lagrangian. The physical mass of the $Z$ is given in
eq. (5.8). The physical electric charge is defined at zero momentum,
then, from eq. (5.10)
\be
e = \et \big(1-\f{z_\gamma}{2}\big)
\ee
The Fermi constant $G_F$, is defined from the $\mu$-decay
process, again at zero momentum. Since the charged current coupling (see
eq. (5.9)) is
modified by a factor
$(1-z_w(1+ M_W^2/M^2)/2)$ and the $W$ mass is given in eq. (5.8)
 we get
\be
\f{G_F}{\sqrt{2}}=\f{\et^2\Big(1-z_w(1+\dd{\f{M_W^2}{M^2}}
)\Big)}{8\st^2{\tilde M}_W^2\Big(1-z_w(1+
\dd{\f{M_W^2}{M^2}})\Big)}
     =\f{e^2}{8\st^2\ct^2 M^2_Z} \Big(1-z_z(1+\f {M_Z^2}{M^2})+z_\gamma\Big)
\ee
where in the second equality we have used eqs. (5.4), (5.8) and (5.12).
Finally, we  define $\s$ and $\c$ by equating the last expression
to the one
in the SM (tree level):
$G_F/\sqrt{2}=e^2/(8\s^2\c^2 M^2_Z)$. We get
\be
\s^2\c^2=\st^2\ct^2\Big(1+z_z(1+\f {M_Z^2}{M^2})-z_\gamma\Big)
\ee
that is
\bea
\s^2 &=& \st^2 \Big( 1+\f{\c^2}{c_{2\theta}}(z_z(1+\f {M_Z^2}{M^2})-
z_\gamma)\Big)\nn\\
\c^2 &=& \ct^2 \Big( 1-\f{\s^2}{c_{2\theta}}(z_z(1+\f {M_Z^2}{M^2})-
z_\gamma)\Big)
\eea

\resection{Calculation of the $\eps$ parameters}

Let us now discuss how the effects of the $\Lt$ and $\Rt$
modify the observables measured at LEP and Tevatron.

Since, in our model, the  modifications due to heavy particles
are contained in the propagators of the standard gauge bosons
(the so-called oblique corrections), we can apply the analysis made in terms
of  the $\eps$ parameters \cite{altarelli}.

Let us start from the $M_W$ measurement. It is customary to  define
\be
\f{M^2_W}{M^2_Z}=
 \c^2\Big[1-\f{\s^2}{c_{2\theta}}\Delta r_W\Big]
\ee
 From the relation ${\tilde M}^2_W=
{\tilde M}^2_Z\ct^2$ we get
\be
\f{M^2_W}{M^2_Z}=
 \c^2\Big[1+
z_z(1+\f{M_Z^2}{M^2})-z_w(1+\f{M_W^2}{M^2})-\f{\s^2}{c_{2\theta}}
(z_\gamma-z_z(1+\f{M_Z^2}{M^2}))\Big]
\ee
so, for comparison
\be
\Delta r_W=z_\gamma+\f{c_{2\theta}}{\s^2}z_w(1+\f{M_W^2}{M^2})
-\f{\c^2}{\s^2}z_z(1+\f{M_Z^2}{M^2})=-X
\ee
where in the second equality we have used eq. (5.5) and
\be
X=2\f{M_Z^2}{M^2}\left(\f g\gs\right)^2
\ee
The neutral current couplings  to the $Z$ are defined by
\be
\L^{neutral}(Z)=-\f{e}{\s\c}\Big(1+\f{\Delta\rho}{2}\Big)Z_\mu\overline\psi
[\gamma^\mu g_V+\gamma^\mu \gamma_5g_A]\psi
\ee
with
\bea
g_V &=& \f{T^3_L}{2}-s^2_{\bar\theta} Q\nn\\
g_A &=& -\f{T^3_L}{2}\nn\\
s^2_{\bar\theta} &=& (1+\Delta k) \s^2
\eea
By using eqs. (5.12) and (5.14) we get
\be
\f{e}{\s\c}=\f{\et}{\st\ct}\Big(1-\f{z_z}{2}(1+\f{M_Z^2}{M^2})\Big)
\ee
For comparison with eq. (5.10), and using eq. (5.5), we obtain
\bea
\Delta\rho &=&
-z_z\f{M_Z^2}{M^2}=-\f{c_\theta^4+s_\theta^4}{c_\theta^2}X\nn\\
\Delta k &=& \f{\c^2}{c_{2\theta}}\Big(z_\gamma-z_z(1+\f{M_Z^2}{M^2})\Big)-
\f{\c}{\s}z_{z\gamma}(1+\f
{M_Z^2}{M^2})=-\f{2c_\theta^2s_\theta^2}{c_{2\theta}}X
\eea

Summarizing, we have the following correspondence between corrections and
observables: $\Delta r_W$ is equivalent to $M_W/M_Z$ which is
measured at Tevatron, $\Delta k$ modifies the
vector coupling $g_V$ and $\Delta\rho$ modifies the neutral coupling
overall strength.
At LEP, $\Delta k$ can be obtained by measuring the forward-backward
 asymmetry at the $Z$
peak. Then, having fixed $\Delta k$,
 $\Delta\rho$ can be determined by the leptonic width.
All these quantities receive contributions also from weak radiative
corrections. In particular they depend quadratically from the top mass
which is still affected by a large error.
 From the point of view of data analysis it turns out to be more convenient
to isolate such contribution in $\Delta\rho$ and define two other linear
combinations which depend only logarithmically on $m_{top}$. They are
the so-called $\eps$ parameters \cite{altarelli}
\bea
\eps_1 &=& \Delta\rho\nn\\
\eps_2 &=& \c^2\Delta\rho+\f{\s^2}{c_{2\theta}}\Delta r_W-2 \s^2\Delta k\nn\\
\eps_3 &=& \c^2\Delta\rho+c_{2\theta}\Delta k
\eea
Using  eqs. (6.3) and (6.8) we get
\bea
\epsilon_1&=&-\f{\c^4+\s^4}{\c^2}~ X\nn\\
\epsilon_2&=&-\c^2~ X\nn\\
\epsilon_3&=&-X
\eea
All these deviations are of order $X$ which contains a double
suppression factor $M_Z^2/M^2$ and $(g/\gs)^2$.
These are the same results one obtains from the definitions of the
$\epsilon_i$
parameters in terms of the self-energies \cite{fri}, \cite{vecax}.
In the $M\to\infty$ limit, the model decouples, as already noticed in sect.
6,
and the $\eps_i$ go to zero. The fact that in the degenerate BESS model
$\eps_3=0$ in this limit, follows
from the $SU(2)_L\otimes SU(2)_R$ custodial symmetry \cite{inami}.

The sum of the SM contributions, functions of the top and Higgs masses,
and the previous deviations has to be compared with the experimental
values for the $\eps$ parameters, determined from the available
LEP data and the $M_W$ measurement from Tevatron \cite{cara}
\bea
\epsilon_1&=&(3.48\pm 1.49)\cdot 10^{-3}\nn\\
\epsilon_2&=&(-5.7\pm 4.19)\cdot 10^{-3}\nn\\
\epsilon_3&=&(3.25\pm 1.40)\cdot 10^{-3}
\label{eps}
\eea
Taking into account the SM values
$(\epsilon_1)_{SM}=4.4\cdot 10^{-3}$,
$(\epsilon_2)_{SM}=-7.1\cdot 10^{-3}$,
$(\epsilon_3)_{SM}=6.5\cdot 10^{-3}$
 for $m_{top}=180~GeV$ and
$m_H=1000~GeV$, we find, from the combinations of the previous
experimental results, the $90\%$ CL limit in the plane $(M, g/\gs)$
given in Fig. 1.
We see that there is
room for relatively light resonances beyond the usual SM spectrum.

\begin{figure}
\epsfysize=8truecm
\centerline{\epsffile{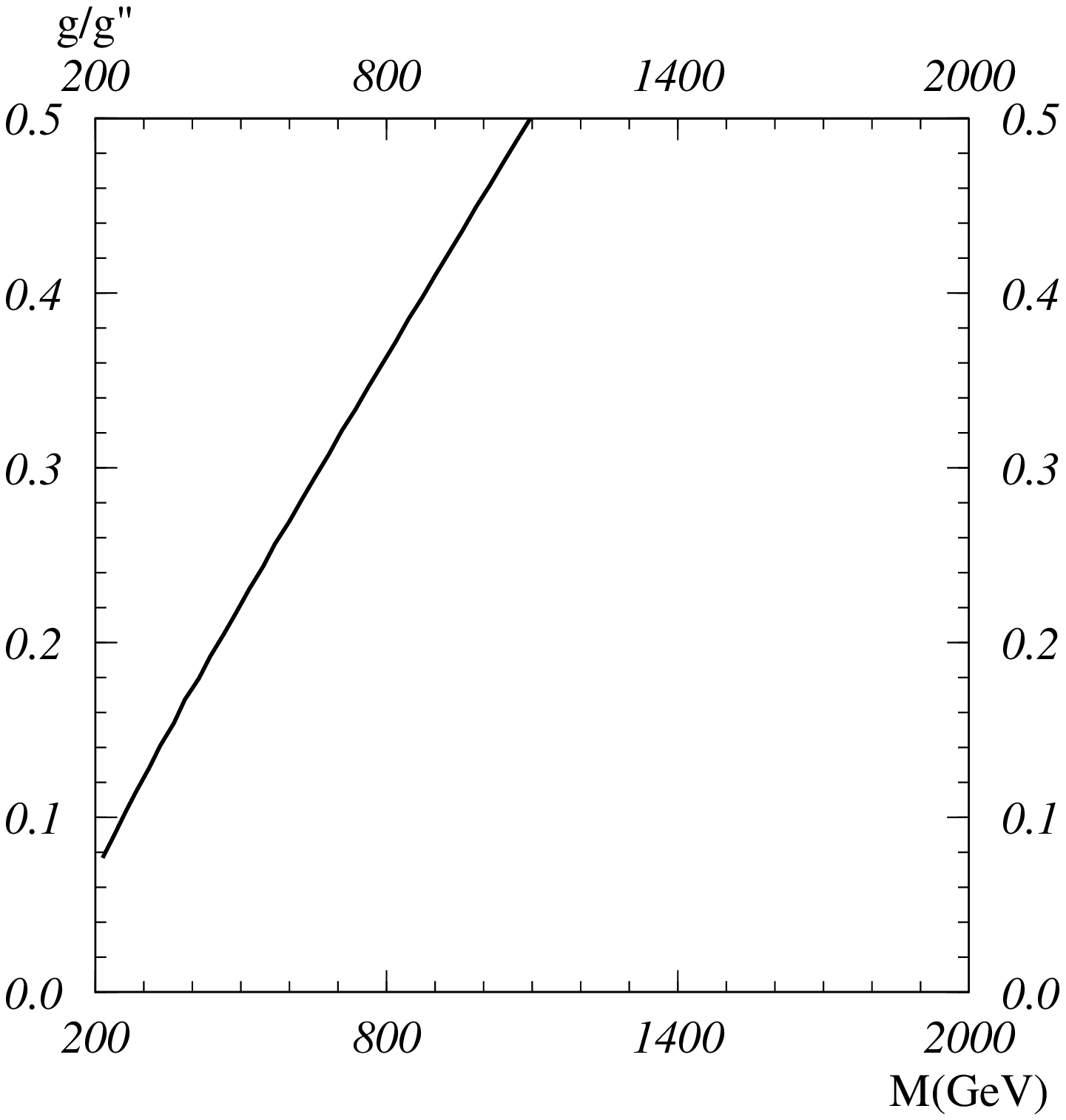}}
\noindent
{\bf Fig. 1} - {\it 90\% C.L. contour on the plane ($M$, $g/g''$) obtained by
comparing the values of the $\epsilon$ parameters from the degenerate BESS
model
with the experimantal data from LEP.
 The allowed region is below the curve.}
\end{figure}

\resection{Anomalous trilinear gauge couplings}

Let us evaluate the anomalous contributions to the trilinear
gauge couplings at the order $q^2/M^2$.
As previously observed, since $\Delta \Rt_\mu$ in eq. (5.2)
does not contain
bilinear and trilinear terms, the elimination of the $\Rt$-field
does not give any
contribution to the anomalous trilinear and quadrilinear terms.

Substituting the solutions (5.1) in eq. (3.3) we get
\newpage

\begin{figure}
\epsfysize=5truecm
\centerline{\epsffile{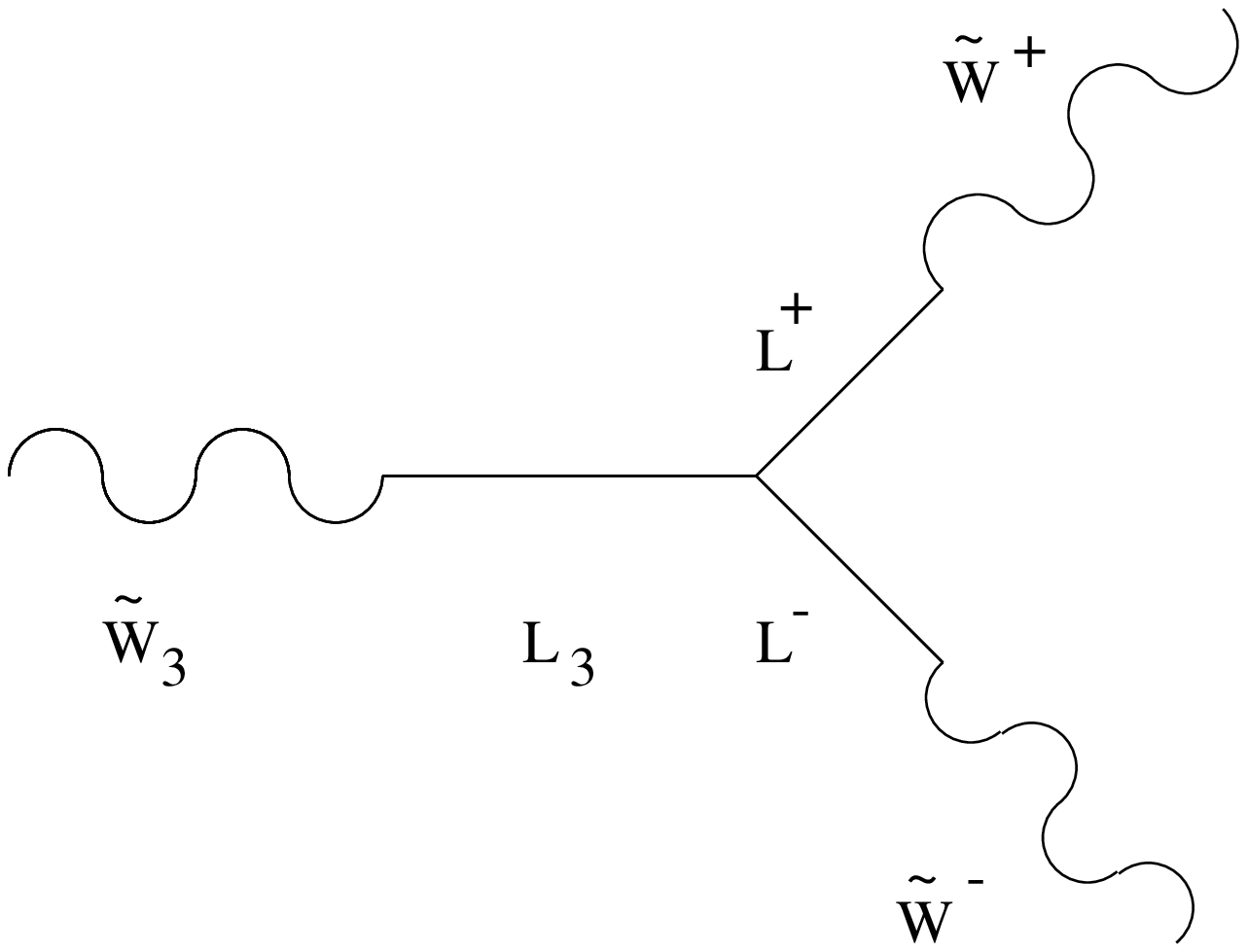}}
\smallskip
\noindent
{\bf Fig. 2} - {\it Feynman diagram contributing to the
anomalous trilinear gauge coupling.}
\end{figure}

\bea
\L^{kin(3)}_{eff}&=&-i\gt\{
(1+z_w)
({\tilde W}^3_{\mu\nu}{\tilde W}^{+\mu}{\tilde W}^{-\nu}+
{\tilde W}^{3\mu}({\tilde W}^-_{\mu\nu}{\tilde W}^{+\nu}-
{\tilde W}^+_{\mu\nu}{\tilde W}^{-\nu}))\nn\\
&-& \f{z_w}{M^2}[
  (\Box_3+\Box_-+\Box_+)(
{\tilde W}^3_{\mu\nu}{\tilde W}^{+\mu}{\tilde W}^{-\nu}+
{\tilde W}^{3\mu}({\tilde W}^-_{\mu\nu}{\tilde W}^{+\nu}-
{\tilde W}^+_{\mu\nu}{\tilde W}^{-\nu}))\nn\\
&+&
((\dnu\dmus {\tilde W}^3_\mu)(\partial_\rho {\tilde W}^{-\nu}){\tilde
W}^{+\rho}
+(\dnu\dmus {\tilde W}^3_\mu)(\dnus {\tilde W}^{+\rho}){\tilde
W}^{-}_\rho\nn\\
&-&
(\dnu\dmus {\tilde W}^-_\mu)(\partial_\rho {\tilde W}^{3\nu}){\tilde
W}^{+\rho}
+(\dnu\dmus {\tilde W}^-_\mu)(\partial^\nu {\tilde W}^{3\rho}){\tilde
W}^{+}_\rho\nn\\
&-&(\dnu\dmus {\tilde W}^-_\mu)(\partial^\nu {\tilde W}^{+\rho}){\tilde
W}^{3}_\rho
+(\dnu\dmus {\tilde W}^-_\mu)(\partial^\rho {\tilde W}^{+\nu}){\tilde
W}^{3}_\rho\nn\\
&-&{\rm h.c.})]\}
\eea
where we have used the notation $\Box_a$ to denote the action of the
D'Alembert operator on the
fields ${\tilde W}^a$, and we have freely integrated by parts.
This result has been
independently checked by evaluating directly the trilinear
${\tilde W}^3{\tilde W}^+{\tilde W}^-$ coupling as coming
from the mixing $\Lt-{\Wt}$ (see Fig.
2) and expanding the propagators of the $\Lt$ fields up to
the order
$q^2/M^2$.

The physical vertices are obviously obtained by substituting
to ${\tilde W}^3$
its expression in terms of the photon and the $Z$ fields and by
performing fields and couplings renormalization according to section 5.

Since the physical process which is relevant for studying
the trilinear
gauge couplings is $e^+e^-\to W^+W^-$, we have
\be
\dmu W^{\pm\mu}=0
\ee
because the final $W$'s are on-shell. Also
\be
\dmu Z^\mu\simeq\dmu A^\mu\simeq 0
\ee
because the $Z$ and the photon  are coupled to light fermions. Therefore we
can
neglect all
the divergences of the fields in eq. (7.1) and we get
\bea
\L^{kin(3)}_{eff}
 &=&ie {\rm ctg}\theta(1+\f {z_z}{2 c_{2\theta}}\f {M_Z^2}{M^2}
-\f {z_w} 2\f {\Box_++M_W^2}{M^2}
-\f {z_w} 2\f {\Box_-+M_W^2}{M^2}
-\f {z_z} 2\f {\Box_Z+M_Z^2}{M^2})\nn\\
&& \Big(Z^{\mu\nu}W^-_\mu W^+_\nu +
       Z^{\nu}(W^-_{\mu\nu}W^{\mu+}-
	W^+_{\mu\nu}W^{\mu -})\Big)\nn\\
&+&ie(1
-\f {z_w} 2\f {\Box_++M_W^2}{M^2}
-\f {z_w} 2\f {\Box_-+M_W^2}{M^2}
+(\f {z_\gamma} 2-z_w)\f {\Box_A}{M^2})\nn\\
&&\Big(A^{\mu\nu}W^-_\mu W^+_\nu +
       A^{\nu}( W^-_{\mu\nu} W^{\mu +}-
	W^+_{\mu\nu}W^{\mu-})\Big)
\eea
 We see
that the tensor structure of this correction
is the same of the trilinear couplings in the SM.

In the study of the reaction
$e^+e^-\to W^+W^-$ at linear colliders, the structure of the
corrections is of the form $(A+B/M^2~s)$, that is we have non trivial form
factors \cite{hag}. However, notice
that the electric charge of the $W$'s as measured by the
coupling with
the photon, turns out to be correct, being
defined at zero transferred momentum and with the $W$'s on shell.

\resection{Masses and eigenstates of spin-1 bosons}

Up to now we have been interested  in the virtual effects only. In the
following we will consider also the possibility of producing directly
the heavy resonances. Therefore we need to keep explicitly the
corresponding fields in the formalism.

By writing  the quadratic part of ${\cal L}$, given in eq. (3.3),
in terms of the charged and the neutral fields  one finds:
\bea
{\cal L}^{(2)} &=& \frac{v^2}{4}[(1+2 a_2)\gt^2 \tW_\mu^+ \tW^{\mu -}+
		      a_2 \gs^2 (\tL_\mu^+ \tL^{\mu -}+\tR_\mu^+
		  \tR^{\mu -})\nn\\
		& &-\sqrt{2}a_2 \gt\gs (\tW_\mu^+ \tL^{\mu -}+\tW_\mu^-
		\tL^{\mu +})]
\nn\\
&+& \frac{v^2}{8}[(1+2 a_2) (\gt^2 \tW_3^2+\gptd \tY^2)+
		      a_2 \gs^2 (\tL_3^2+\tR_3^2)\nn\\
		& &- 2 \gt \gpt \tW_{3\mu}\tY^\mu
		-2 \sqrt{2}a_2\gs (\gt \tW_3 \tL_3^\mu +\gpt \tY_\mu
		 \tR_3^\mu)]
\eea
The reason to introduce $\tilde\Lt$ and $\tilde\Rt$ is to distinguish
them from the mass eigenstates.

In the charged sector the fields $R^\pm$ are unmixed
 for any value of $\gs$.
Their mass is given by:
\be
M^2_{R^\pm}=\f{v^2}{4} a_2 \gs^2\equiv M^2
\ee
We will parameterize the model in terms of $\gs$ and $M$.

The mass matrix in the charged sector $(\tW,\tL)$ is
\be
M^2_{charged} =\f{v^2} 4 \gt^2\left(
\begin{array}{cc}\displaystyle{
1+2\f {x^2} r} & \dd{-\sqrt{2} \f x r}  \\
\dd{-\sqrt{2}\f x r} &\dd{ \f 1 r}
\end{array}
\right)
\ee
where
\be
x=\f {{\tilde g}} \gs,~~~~~r=\f{v^2} 4\f{\gt^2}{M^2}
\ee
At the order $x^2$ the eigenvalues are
\be
M^2_{{W}^\pm}=\f{v^2}{4} \gt^2[1-2\f{x^2}{1-r}+\cdots]~~~,~~~~~
M^2_{{L}^\pm}=\f{v^2}{4} \f{\gt^2} r [1+2 \f{x^2} {1-r}+\cdots]
\ee
Let us call ${\bf C}$ the matrix which transforms the fields appearing
in the lagrangian (8.1) into the charged eigenstates. We have
\be
\left(
\begin{array}{c}W^\pm\\L^\pm\end{array}\right)=
{\bf C}^{-1}\left(\begin{array}{c}\tW^\pm\\\tL^\pm\end{array}\right)=
\left(
\begin{array}{cc}
\cos\phi & \sin\phi \\
-\sin\phi & \cos\phi\end{array}\right)
\left(\begin{array}{c}\tW^\pm\\\tL^\pm\end{array}\right)
\ee
where
\be
\tan\phi=\f k {1+\sqrt{1+k^2}},~~~~
k=\f{2\sqrt{2}x}{1-r-2 x^2}
\ee
The physical particle ${L}^\pm$ are a combination of $\tL^\pm$
and $\tW^\pm$, which, for
small values of $x$, and for $M\to\infty$ ($r\to 0$),
are mainly oriented along the $\tL^\pm$
direction,
as it can be seen from the following relations
\bea
W^\pm&=&\left(1-x^2\right)\tW^\pm+\sqrt{2}x
\tL^\pm\nn\\
L^\pm&=&-\sqrt{2} x \tW^\pm+\left(1-x^2\right)\tL^\pm
\eea

In the neutral sector there is, as expected, a strictly massless combination
which corresponds to the physical photon associated to the unbroken
$U(1)$ gauge group. As already discussed in ref. \cite{assiali} we
perform the substitution
\bea
{\hat W}^3&=& \ct \tW^3-\st \tY\nn\\
{\hat Y}&=& \st \tW^3+\ct \tY
\eea
The neutral part of the lagrangian (8.1)
is then given by
\bea
{\cal L}^{(2)}_{neutral} &=&
\frac{v^2}{8}[ G^2 \hat W_3^2+a_2(\f{\gt^2-\gptd}{G} \hat W_3
+2\f{\gt\gpt}{G}\hat Y-\f{\gs}{\sqrt{2}}(\tL_3+\tR_3))^2
\nn\\&+&
a_2(G \hat W_3
-\f{\gs}{\sqrt{2}}(\tL_3-\tR_3))^2]
\eea
where $G=\sqrt{\gt^2+\gptd}$.
Inspecting this expression, it is natural to define the new linear
combinations
\bea
 \gamma &=& \hat Y\cos\psi+  \f{\tL_3+\tR_3}{\sqrt{2}}\sin\psi\nn\\
\hat V_3 &=& -\hat Y\sin\psi+
  \f{\tL_3+\tR_3}{\sqrt{2}}\cos\psi
\eea
where
\be
\tan\psi=2\st x
\ee
We then finally obtain
\bea
{\cal L}^{(2)}_{neutral} &=&
\frac{v^2}{8}[ G^2 \hat W_3^2+\f {E_V^2} {R_V}(G \hat W_3
-\f{G}{E_V}\hat V_3)^2
\nn\\&+&
\f {E_A^2}{R}(G \hat W_3
+\f G{E_A} \hat A_3)^2]
\eea
where
\bea
R&=&\f r {\ct^2}\nn\\
R_V&=&\f R {1+4 x^2\st^2}\nn\\
E_A&=&\f x{\ct}\nn\\
E_V&=&\f{\tilde\cdt}{\ct}\f{x}{\sqrt{1+4 x^2 \st^2}}\nn\\
\hat A_3&=&\f {\tR_3-\tL_3}{\sqrt{2}}
\eea
The mass matrix for the neutral case in the basis $(\hat W_3,\hat
V_3,\hat A_3)$ is the following
\be
M^2_{neutral} =\f{v^2} 4 G^2\left(
\begin{array}{ccc}\displaystyle{
1+\f {E_V^2} {R_V}+\f{E_A^2} R} & \dd{- \f {E_V}{R_V}} &
\dd{ \f{E_A}{R}} \\
\dd{-\f {E_V} {R_V}} &\dd{ \f 1 {R_V}} & 0\\
 \dd{\f {E_A} {R}} &0&\dd{ \f 1 {R}}
\end{array}
\right)
\ee
At the order $x^2$ the eigenvalues are
\bea
M^2_{Z}&=&\f {v^2} 4 G^2 \left(1-
\f {1+\tilde c_{2\theta}^2}{\ct^2-r}x^2+\cdots\right)\nn\\
M^2_{L_3}&=&\f {v^2} 4 \f{G^2} R\left(1+\f{1-2
r\st^2+\sqrt{\tilde c_{2\theta}^2+ 4 r^2\st^4}}{\ct^2-r}x^2+
\cdots\right)\nn\\
M^2_{R_3}&=&\f {v^2} 4 \f{G^2} R\left(1+\f{1-2
r\st^2-\sqrt{\tilde c_{2\theta}^2+ 4 r^2\st^4}}{\ct^2-r}x^2+\cdots\right)
\eea
The relation between the $hat$ fields and the mass eigenstates is given by
\be
\left(
\begin{array}{c}
\hat W_3\\ \hat V_3 \\ \hat A_3
\end{array}
\right)=
V
\left(
\begin{array}{c}
Z\\ L_3 \\ R_3
\end{array}
\right)
\ee
with
\be
V=
\left(
\begin{array}{ccc}\displaystyle{
N_1} &
\dd{N_2\f  {1-\lambda_2 R_V}{E_V}}
&
\dd{-N_3\f  {1-\lambda_3 R}{E_A}}
 \\
&&\\
\dd{N_1 \f {E_V}{1-R_V\lambda_1}} &\dd{ N_2} &
 \dd{-N_3\f{E_V}{E_A}
\f{1-\lambda_3 R}{1-\lambda_3 R_V}}\\
&&\\
\dd{-N_1 \f{E_A}{1-R\lambda_1}}  &
\dd{-N_2\f{E_A}{E_V}
\f{1-\lambda_2 R_V}{1-\lambda_2 R}}
 & N_3
\end{array}
\right)
\ee
where
\be
\lambda_1=4 \f{ M^2_Z}{v^2 G^2},~~~~~
\lambda_2=4  \f{ M^2_{L_3}}{v^2 G^2},~~~~
\lambda_3=4  \f{ M^2_{R_3}}{v^2 G^2}
\ee
and
\bea
N_1 &=& \left(1+\f{E_V^2}{(1-R_V \lambda_1)^2}+
\f{E_A^2}{(1-R \lambda_1)^2}\right)^{-1/2}\nn\\
N_2 &=& \left(1+\f{(1-R_V \lambda_2)^2}{E_V^2}(1+
\f{E_A^2}{(1-R \lambda_2)^2})\right)^{-1/2}\nn\\
N_3 &=& \left(1+\f{(1-R\lambda_3)^2}{E_A^2}(1+
\f{E_V^2}{(1-R_V \lambda_3)^2})\right)^{-1/2}
\eea
Let us call ${\bf N}$ the matrix which transforms the fields
appearing in the lagrangian (8.1) into the neutral eigenstates
\be
\left(
\begin{array}{c}\gamma \\ Z \\ L_3 \\ R_3\end{array}\right)=
{\bf N}^{-1}
\left(
\begin{array}{c}\tY \\ \tW_3 \\ \tL_3 \\ \tR_3\end{array}\right)
\ee
In the limit
of $r\to 0$ and small $x$, by
retaining only the first order in $x$, we get
\be
\left(
\begin{array}{c}\gamma \\ Z \\ L_3 \\ R_3\end{array}\right)
\simeq
\left(\begin{array}{cccc}
 \ct &\st &\sqrt{2} \st x &\sqrt{2} \st x \\
-\st & \ct &\sqrt{2}\ct x &\dd{-\sqrt{2}\f{\st^2}{\ct}x} \\
0 &-\sqrt{2} x & 1 & 0 \\
\dd{- \sqrt{2} x \f{\st}{\ct}} &0&0&1\end{array}\right)
\left(
\begin{array}{c}\tY \\ \tW_3 \\ \tL_3 \\ \tR_3\end{array}\right)
\ee
where we have used the eqs.  (8.9), (8.11), (8.14) and (8.18).
We see that ${Z}$, ${L}_3$ and ${R}_3$ are
essentially aligned along the combinations $(\ct~\tilde W_3-\st~ \tilde Y)$,
$\tL_3$ and $\tR_3$ respectively.
Unlike the charged case, however,
the physical state $R_3$ is not completely decoupled, in fact at the
leading order,
it possesses a tiny component along the
$\tilde Y$ direction. The $L_3$ state has in turn a small contribution from
the
$\tilde W_3$ field.

\resection{Fermionic couplings}

 From the charged part of the fermionic lagrangian given in eq. (3.6), by
using the relations (8.6), we can read directly the couplings to the
fermions
\be
{\cal L}_{charged}=
-\left(a_W W_\mu^-+a_L L_\mu^-\right)J_L^{(+)\mu}+{\rm h.c.}
\ee
where
\bea
a_W&=&\f \gt{\sqrt{2}}C_{11}\nn\\
a_L&=&\f \gt{\sqrt{2}}C_{12}
\eea
where $ C_{ij}$ are the matrix elements of the matrix ${\bf C}$
defined in eq. (8.6), and
\be
J_L^{(\pm)\mu}=\bar\psi_L\gamma^\mu\tau^{(\pm)}\psi_L
\ee
with
\be
\tau^{(\pm)}=\f{\tau_1\pm i \tau_2} 2
\ee
Let us notice that the $R^{\pm}$ are not coupled to the fermions. In fact
one can easily check that they have no mixing whatsoever and therefore these
states will be absolutely stable as ensured by the phase invariance
$R^{\pm}\to\exp(\pm i\alpha)R^\pm$.

For the neutral part we get
\bea
{\cal L}_{neutral}&=& -\Big\{
\left[A J_L^{(3)\mu}+BJ_{em}^\mu \right]Z_\mu+
\left[C J_L^{(3)\mu}+DJ_{em}^\mu \right]L_{3\mu}+
\left[E J_L^{(3)\mu}+FJ_{em}^\mu \right]R_{3\mu}\nn\\&+&
eJ_{em}^\mu\gamma_\mu \Big\}
\eea
where
\be
e=\gt\st\cos\psi
\ee
and
\be
\begin{array}{lll}
A=G V_{11}&&B=\dd{-G\st^2(V_{11}+\f{\ct}{\st}\sin\psi V_{21})}\\\\
C=G V_{12}&&D=-\dd{G\st^2(V_{12}+\f{\ct}{\st}\sin\psi V_{22})}\\\\
E=G V_{13}&&F=\dd{-G\st^2(V_{13}+\f{\ct}{\st}\sin\psi V_{23})}
\end{array}
\ee
with $V_{ij}$ are the matrix element of the matrix $V$
given in eq.
(8.18). In the usual limit we get, at the order $x^2$,
\be
\begin{array}{lll}
A\simeq G (1-\dd{\f{\st^4+\ct^4}{\ct^2}}x^2)
&&B\simeq \dd{-G\st^2(1-\f{1-2\ct^4}{\ct^2}x^2)}\\\\
C\simeq-\sqrt{2}G\ct x &&D\simeq 0\\\\
E\simeq \dd{\sqrt{2}G\f{\st^2}{\ct}x}&&F\simeq
\dd{-\sqrt{2}G\f{\st^2}{\ct}x}
\end{array}
\ee

\resection{Trilinear couplings}

Starting from the original trilinear couplings given in eq. (3.4) and
using the relations (8.6) and (8.21), we can evaluate all the trilinear
couplings among the physical particles of the model. We get
\bea
\L^{kin~(3)} &=&
 i \sum_{a,i,j} g_{O_a V^+_i V^-_j}\Big[O_a^{\mu\nu}V_{i\mu}^- V_{j\nu}^+ +
       O_a^{\nu}(V^-_{i\mu\nu}V_{j\mu}^+ -
	V^+_{i\mu\nu}V^-_{j\mu})\Big]\nn\\
& &+i\sum_{a} g_{O_a R^+ R^-} [O_a^{\mu\nu}R^-_\mu R^+_\nu +
       O_a^{\nu}( R^-_{\mu\nu} R^+_\mu-
	R^+_{\mu\nu}R^-_\mu)\Big]
\eea
 where $O_a=\gamma,~Z,~L_3,~R_3$ $(a=1,2,3,4)$,
 $V_i^\pm= W^\pm,~L^\pm$ $(i=1,2)$ and
 \bea
g_{O_a V^+_i V^-_j} &=&
\gs (x C_{1i} C_{1j} N_{2a}+ \f 1{\sqrt{2}}C_{2i} C_{2j} N_{3a})\\
g_{O_a R^+ R^-} &=&
\gs \f 1{\sqrt{2}} N_{4a}
\eea
As we noticed in the previous section,
there are no mixings of the $\tR^{\pm}$ with  $\tW^{\pm}$ and
$\tL^{\pm}$ (see eq. (8.1)), as a consequence
there are no trilinear couplings involving a single charged $R$ particle.
Furthermore it is easy to check that
 $g_{\gamma W^+L^-}=0$.

\resection{Renormalization procedure}

To identify the physical quantities in our model we proceed
in the same way as in sect. 5.
We again choose as input parameters the following
physical constants: the electric charge, the mass of the $Z$
and the Fermi constant.

Concerning the Fermi constant there is a general proof
in ref. \cite{assiali} stating that its relation with $v^2$
is the same of the SM one.
Let us summarize the principal steps of the proof.
Keeping into account the $W^\pm$ and $L^\pm$ exchanges in the
$\mu$-decay
process,
the Fermi constant is given by
\be
\f{G_F}{\sqrt{2}}=\f{\gt^2} 8 \left(\f{C_{11}^2}{M^2_{W^\pm}}+
\f{C_{12}^2}{M^2_{L^\pm}}\right)=\f {\gt^2} 8 (M^2_{charged})^{-1}_{11}
=\f 1 {2v^2}
\ee
where $C_{ij}$ are given in eq. (8.6) and $M^2_{charged}$
in (8.3). The second equality in the previous equation follows
from the relation
\be
{\bf C} (M_D^2)^{-1} {\bf C}^{-1}= (M^2_{charged})^{-1}
\ee
where $M_D^2$ is the diagonal form  of the charged mass matrix.
Recalling that the electric charge and the mass of the $Z$ are given
by
\bea
e &=& \gt \st \cos\psi \\
M^2_Z &=& {\tilde M}^2_Z  \lambda_1
\eea
with
\be
{\tilde M}^2_Z=\f 1 4 v^2 \f{\gt^2}{\ct^2}
\ee
and $\lambda_1$ is obtained by the diagonalization of the matrix in eq.
(8.15).

By using the previous equations we can write the following relation
($\alpha=e^2/4\pi$)
\be
\ct^2\st^2=\f{\pi\alpha}{\sqrt{2}G_F} \f 1{M_Z^2}\f {\lambda_1}
{\cos^2\psi}
\ee
from which
\be
\ct^2=\f 1 2 +\sqrt{ \f 1 4 - \f{\pi\alpha}{\sqrt{2}G_F} \f 1{M_Z^2}\f
{\lambda_1}
{\cos^2\psi}}
\ee
 From eq. (8.12) and using eq. (11.3) we obtain
\be
\sin\psi=2\f e \gs
\ee
from which
\be
\gt =\f e {\st\sqrt{1-\dd{\f{ 4 e^2}{\gs^2}}}}
\ee
Let us notice that eq. (11.7), after using (11.9), involves only the angle
$\tilde\theta$ and the measured quantities. Solving this equation in
$\tilde\theta$ we can obtain $\gt$  with the help of eq. (11.9).
In this way all the original parameters of the lagrangian
are expressed in terms of the observed quantities.
In the numerical work we solve the equation (11.7) by
iteration taking advantage of the fact that in the limit
$\gs\to\infty$ we get back the SM.
One can also solve the equation perturbatively in $x$.
It is easy to verify that, at the order $x^2$ and at the
first order in $r$ (see eq. (8.4)), the procedure coincides
with the one of sect. 5, and one recovers
the equation (5.15). In fact this follows immediately from
the expansion of $\lambda_1$ (see eqs. (8.19) and (8.16))
and $\cos^2\psi$ at the same order
\be
\lambda_1\simeq 1-z_z(1+\f{M^2_Z}{M^2})~~~~~~\cos^2\psi\simeq 1-z_\gamma
\ee

\resection{Widths}

In the neutral sector the couplings of the
 fermions to the gauge bosons are
\be
-\f 1 2 \bar \psi [(v_Z^f+\gamma_5 a_Z^f)\gamma_\mu Z^\mu+
(v_{L_3}^f+\gamma_5 a_{L_3}^f)\gamma_\mu L_3^\mu+
(v_{R_3}^f+\gamma_5 a_{R_3}^f)\gamma_\mu R_3^\mu]\psi
\ee
where $v^f$ and $a^f$ are the vector and axial vector
couplings given by
\bea
v_Z^f &=&   A T_3^L + 2B Q_{em}\nn\\
a_Z^f&=& A T_3^L\nn\\
v_{L_3}^f &=&   C T_3^L + 2D Q_{em}\nn\\
a_{L_3}^f&=& C T_3^L\nn\\
v_{R_3}^f &=&   E T_3^L + 2F Q_{em}\nn\\
a_{R_3}^f&=& E T_3^L
\eea
with $A,~B,~C,~D,~E,~F$ given in eq. (9.7).
The total width of a vector boson $V$ corresponding to
the decay into fermion-antifermion is
\be
\Gamma^{fermion}_V= \Gamma_V^h+3(\Gamma_V^l+\Gamma_V^\nu)
\ee
where $\Gamma_V^h$ includes the contribution of all the
allowed quark-antiquark decays.
The partial widths are given by
\be
\Gamma_V^f=\f  {M_V}{48\pi}
F(m_f^2/M^2_V)
\ee
with
\be
F(r_f)= (1-4 r_f)^{1/2} ((v_V^f)^2(1+2 r_f) +(a_V^f)^2(1-4 r_f))
\ee
and $m_f$ the mass of the fermion.

The other possible decay channel for a neutral vector boson
 $V$ is the one corresponding to
the decay into a $WW$ pair. The  partial width is
\bea
\Gamma_V^W &=&\f{M_V}{192 \pi} g^2_{VW^+W^-}
\left(1-4 \f{M^2_W}{M^2_V}\right)^{3/2}\left(\f{M_V}{M_W}
\right)^4\nn\\
&\times & \left[ 1 + 20 \left(\f{M_W}{M_V}
\right)^2 +12 \left(\f{M_W}{M_V}
\right)^4\right]
\eea

Concerning the charged resonances, only the $L^\pm$
decay into fermions (see sect. 9).
The leptonic width neglecting the fermionic mass corrections, is
\be
\Gamma(L^-\to l\bar\nu_l)= \f 1 {24\pi} a_L^2 M_{L} \equiv \Gamma_L^0
\ee
with $a_{L}$ given in eq. (9.2).

The decays into quark pairs are given by
\be
\Gamma(L^-\to q'\bar q)= 3 |V_{qq'}|^2 \Gamma_L^0
\ee
where $V_{qq'}$ are the relevant Kobayashi-Maskawa matrix elements.
In the case of the $ b \bar t $
decay, we have taken into account the correction from the mass of the top
\be
\Gamma(L^-\to b\bar t)= 3 |V_{tb}|^2 (1-\f 3 2 r_t + \f 1 2 r_t^3)\Gamma_L^0
\ee
where $r_t=m_t^2/M_L^2$.

The other possible decay channel for  $L^{\pm}$
is the one corresponding to
the decay into a $WZ$ pair. The  partial width is
\bea
\Gamma_L^{WZ} &=&\f{M_{L}}{192 \pi} g^2_{ZW^+L^-}
\left[\left(1- \f {M_Z^2-M_W^2}{M_L^2}\right)^2
-4\f{M^2_W}{M^2_L}\right]^{3/2}\left(\f{M_L^4}{M_W^2 M^2_Z}
\right)\nn\\
&\times & \left[ 1 + 10 \left(\f{M_W^2 +M_Z^2}{M_L^2}
\right) + \f{M_W^4+M_Z^4+10 M_W^2 M_Z^2}{M_L^4}
\right]
\eea

Let us now give the previous formulae for the widths in the $\gs\to\infty$
limit
(at the order
$(g/\gs)^2$) and neglecting the
 mass corrections. For the fermionic channel we get
\bea
\Gamma(L_3\to e^+e^-)&=&\f{\sqrt{2}G_FM_W^2}{12\pi}M_{L_3}
\left(\f g \gs\right)^2\nn\\
\Gamma(R_3\to e^+e^-)&=&\f{5\sqrt{2}G_FM_W^2}{12\pi}
\f{\s^4}{\c^4}
M_{R_3}\left(\f g \gs\right)^2\nn\\
\Gamma(L^-\to e\bar\nu)&=&\f{\sqrt{2}G_FM_W^2}{6\pi}M_{L}
\left(\f g \gs\right)^2
\eea
and the total fermionic widths are
\bea
\Gamma^{fermion}_{L_3}&=&\f{2\sqrt{2}G_FM_W^2}{\pi}M_{L_3}
\left(\f g \gs\right)^2\nn\\
\Gamma^{fermion}_{R_3}&=&\f{10\sqrt{2}G_FM_W^2}{3\pi}\f {\s^4}
{\c^4}M_{R_3}\left(\f g \gs\right)^2\nn\\
\Gamma^{fermion}_{L^\pm}&=&\f{2\sqrt{2}G_FM_W^2}{\pi}M_{L^\pm}
\left(\f g \gs\right)^2
\eea
By reading the relevant trilinear gauge
couplings from
 eq. (10.2), and performing the same limit we have (with $r$ defined in
(8.4))
\bea
g_{L_3W^+W^-} &\simeq & \sqrt{2} \gt~ r~ \f g \gs\nn\\
g_{R_3W^+W^-} &\simeq & \sqrt{2} \f{\st^2}{\ct^2}\gt~ r~ \f g \gs\nn\\
g_{ZW^+L^-} &\simeq&  \sqrt{2} \f{\gt}{\ct}~ r~ \f g \gs
\eea
Then, substituting in eqs. (12.6) and (12.10),
we get
\bea
\Gamma^{WW}_{L_3}&=&\f{\sqrt{2}G_FM_W^2}{24\pi}M_{L_3}
\left(\f g \gs\right)^2\nn\\
\Gamma^{WW}_{R_3}&=&\f{\sqrt{2}G_FM_W^2}{24\pi}\f{\s^4}{\c^4}
M_{R_3}\left(\f g \gs\right)^2\nn\\
\Gamma^{WZ}_{L^\pm}&=&\f{\sqrt{2}G_FM_W^2}{24\pi}M_{L^\pm}
\left(\f g \gs\right)^2
\eea
It may be useful to compare the widths of $\lmu$ and $\rmu$ into
vector boson pairs with those into fermions:
\bea
\Gamma_{L_3}^{fermion}&=&48~\Gamma_{L_3}^{WW} \nn\\
\Gamma_{R_3}^{fermion}&=&80~\Gamma_{R_3}^{WW} \nn\\
\Gamma_{L^\pm}^{fermion}&=&48~\Gamma_{L^\pm}^{WZ}
\eea
We see that the total fermionic channel is dominant due to the multiplicity.

We conclude this section with some remarks about the decay
of the vector mesons $\lmu$ and $\rmu$.
In the present, effective, description of the electroweak symmetry-breaking,
the Goldstone bosons described by the field  $U$ given in eq. (\ref{lg1})
become unphysical scalars eaten
up by the ordinary gauge vector bosons $W$ and $Z$.
The absence of couplings among $U$ and the states $L$ and $R$
results in a suppression of the decay rate of these
states into $W$ and $Z$. Consider, for instance, the decay of
the new neutral gauge bosons into a $W$ pair. In a model with
only vector resonances this decay channel is largely the dominant one.
The corresponding width is indeed given by \cite{pierre43}
\be
\Gamma (V_0\to WW)=\f{\sqrt{2}G_F}{192\pi} \frac{M^5}{M_W^2}
\left(\f g \gs\right)^2
\ee
and it is enhanced with respect to the partial width into a fermion
pair, by a factor $(M/M_W)^4$ \cite{pierre43}
\be
\Gamma(V_0\to {\bar f} f)\approx G_F M_W^2 \left(\frac{g}{\gs}\right)^2 M
\ee
This fact is closely related to the existence of a coupling of order $\gs$
among $V_0$ and the unphysical scalars absorbed by the $W$ boson.
Indeed the fictitious width of $V_0$ into these scalars provides,
via the equivalence theorem \cite{equ}, a good approximation to
the width of $V_0$ into a pair of longitudinal $W$ and it is
precisely given by eq. (12.16).

On the contrary, if there is no direct coupling among the new gauge bosons
and
the would-be Goldstone bosons which provide the longitudinal
degree of freedom to the $W$, then their partial width
into longitudinal $W$'s will be suppressed compared to the
leading behaviour in eq. (12.16), and the width
into a $W$ pair could be similar to the fermionic width.
In fact, as we have explicitly checked (see eq. (12.13)) the trilinear
couplings
between the new gauge bosons and a $W$ pair
is no longer of order $ (g/\gs)$, but of the order $ (g/\gs)~r$.
The same argument also holds for the charged case.

Numerically, the comparison between the degenerate case (D-model)
and the
BESS model with only vector resonances (V-model)
is shown in Table I, for a choice of the parameters of
the model ($M_{L^+}=1~TeV$, $g''=13$
 and no direct coupling of $L^+$ to
fermions). The V-model features an enhancement of the $WZ$
channel, common to the usual strong interacting models. The D-model has no
such an enhancement.

\begin{table}
\begin{center}
\begin{tabular}{l c c c c}
\hline \hline
& & & & \\
{\rm D-model} & $\Gamma_{L^+}~ (GeV)$ & Br($L^+ \to e\nu$) &
Br($L^+ \to ud$) & Br($L^+ \to WZ$) \\
 & 0.18 & $8.1 \times 10^{-2}$ & $2.4 \times 10^{-1}$ &
$2.2 \times 10^{-2}$ \\ \hline
\hline
& & & & \\
{\rm V-model} & $\Gamma_{V^+}~(GeV)$ & Br($V^+ \to e\nu$)
& Br($V^+ \to ud$) & Br($V^+ \to WZ$) \\
 & 12.1 & $5.9 \times 10^{-4}$ & $1.7 \times
10^{-3}$ & $9.9 \times 10^{-1}$ \\ \hline
\hline
\end{tabular}
\end{center}
\noindent
 {\bf Table I}: {\it Comparison between the degenerate BESS model (D-model)
and the
BESS model with only vector resonances (V-model)
for the total width and branching ratios of $L^+$
(D-model) and $V^+$ (V-model) with the choice of
parameters: $M=1~TeV$, $g''=13$
and no direct coupling of $L^+/V^+$ to fermions.}%
\end{table}

As already noticed, in usual strong interacting
models an enhancement of $W_L W_L$ scattering is expected. Due to the
previous
considerations, our case is quite different. If we study $W_L W_L$ scattering
the lowest order result violates unitarity at energies above 1.7 $TeV$, as in
the
standard model in the formal limit $m_H \to \infty$. So we expect our model
to be valid only up to energies of this order.

\resection{Degenerate BESS at Tevatron}

Data from the Fermilab Tevatron Collider, collected by the CDF collaboration
\cite{cdf} establish limits on the model parameter space. Their search was
done
through the decay $W' \to e \nu$, assuming standard couplings of the $W'$ to
the fermions. Their result can be easily translated into a limit for the
degenerate
BESS model parameter space.
\begin{figure}
\epsfysize=8truecm
\centerline{\epsffile{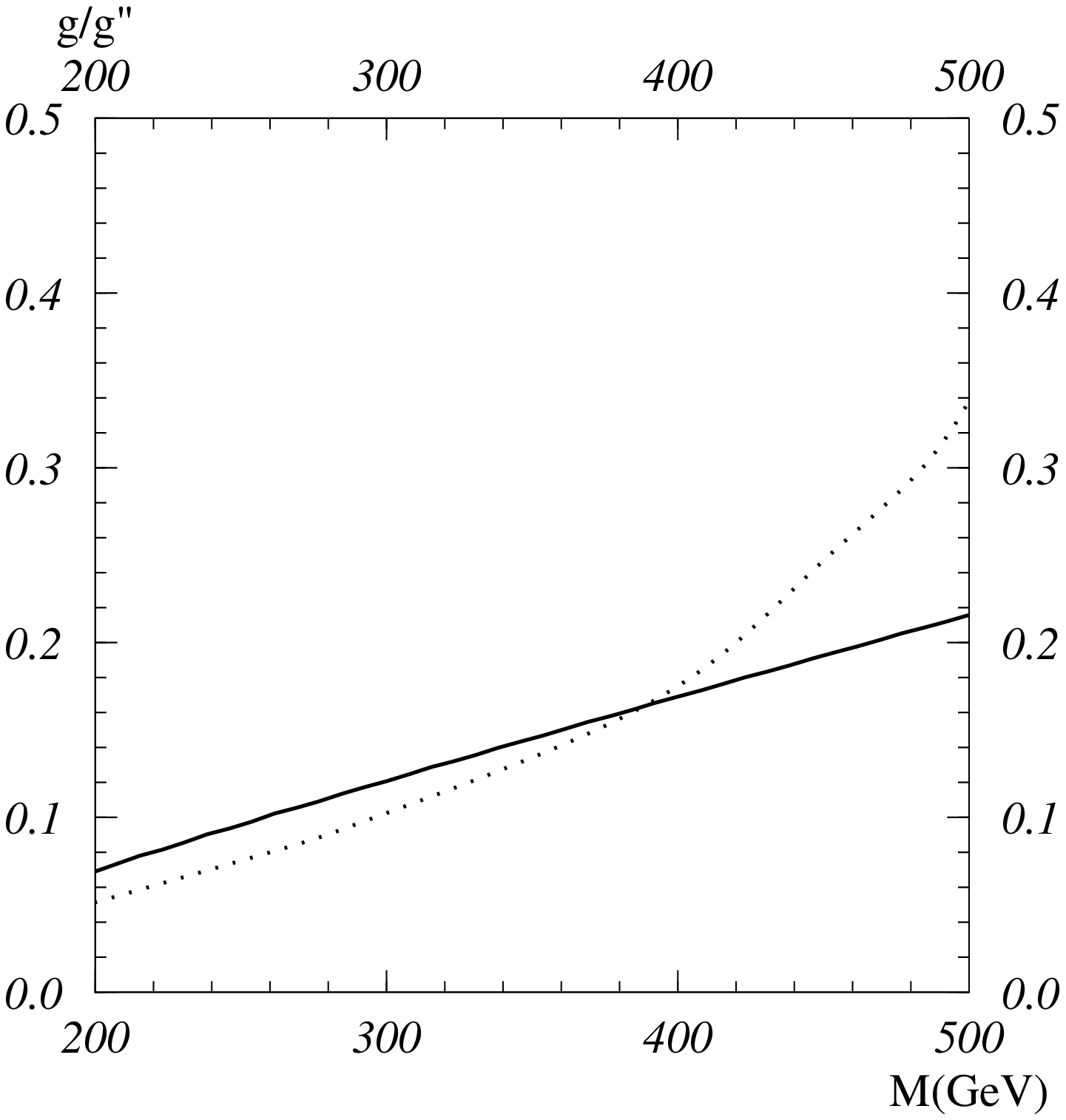}}
\noindent
{\bf Fig. 3} - {\it 95\% C.L. upper bounds on $g/g''$ vs. $M$ from LEP data
(continuous line) and CDF (dotted line).
LEP bounds are obtained from  the $\epsilon$ parameters, while
CDF limits come from $\sigma B (p\bar p \to \ell \nu )$ at $\sqrt{s}=1.8~TeV$
and with an integrated luminosity of $19.7 pb^{-1}$.}
\end{figure}

In Fig. 3 these limits are
shown in terms of the mass of the $R^\pm$ resonance
(equivalent to $M$, see eq. (8.2))
 and the ratio of
coupling constants $g/g''$. Actually the parameter of the model is $x$
as given in eq.
(8.4). Here, for simplicity of notation, we call it $g/g''$.
The limit from CDF (dotted line) is compared with the result obtained from
LEP (continuous line). The excluded region is above the two curves.
The figure was obtained using the CDF 95\% C.L. limit
on the $W'$ cross-section times the branching ratio and comparing this limit
with the predictions of our model at fixed $g/g''$, thus
giving a limit for the $R^\pm$
mass. This procedure was then iterated for various values of $g/g''$.
The statistical significance of the plot is that of a 95\% C.L. limit in one
variable, the mass, at a given value of $g/g''$. The limit from CDF is more
restrictive for low resonance masses, while LEP limit is more restrictive for
higher mass values.

\resection{Degenerate BESS at $e^+e^-$ colliders}

We have considered the sensitivity of the model at LEP2 and future $e^+e^-$
linear colliders, for different options of total centre of mass energies and
luminosities.

We have analyzed cross-sections and asymmetries for the channel
$e^+e^-\rightarrow f^+f^-$ and $e^+e^-\rightarrow W^+W^-$ in the Standard
Model and in the degenerate BESS model at tree level.
The BESS states relevant for the
analysis at $e^+e^-$ colliders are $L_3$ and $R_3$. The two vector bosons are
degenerate in mass in the large $g''$ limit. The $L_3$ mass is larger than
the $R_3$ mass due to terms of the order $(g/g'')^2$ and higher (see eq.
(8.16)).

If the masses of the resonances are below and not very far from
 the collider energy,
due to beamstrahlung and
synchrotron radiation, in a high energy collider, one expects
to see two very close narrow peaks below the maximum c.m. energy even
without having to tune the beam energies. However in this paper we do not
consider the direct production of $R_3$ and $L_3$ from $e^+e^-$.
If instead the masses are higher than the maximum c.m. energy, they will
give rise to indirect effects in the $e^+e^-\rightarrow f^+f^-$ and
$e^+e^-\rightarrow W^+W^-$ cross-sections, which we discuss below.

For the purposes of our calculation we have assumed that it will
be possible to separate $e^+e^-\rightarrow W^+_L W^-_L$, $e^+e^-\rightarrow
W^+_L W^-_T$, and $e^+e^-\rightarrow W^+_T W^-_T$. A similar analysis
for the BESS model with only vector resonances was given in ref.
\cite{pierre43}.

In the fermion channel our study is based on the following observables:
\bea
&\sigma^{\mu},~~\sigma^h\nn\\
&A_{FB}^{e^+e^- \to \mu^+ \mu^-},~~ A_{FB}^{e^+e^- \to {\bar b} b}\nn\\
&A_{LR}^{e^+e^- \to \mu^+ \mu^-},~~A_{LR}^{e^+e^- \to {\bar b} b},~~
A_{LR}^{e^+e^- \to {had}}\nn\\
\eea
where $A_{FB}$ and $A_{LR}$ are the forward-backward and left-right
asymmetries, and $\sigma^{h(\mu)}$ is the total hadronic ($\mu^+\mu^-$)
cross-section.

The total cross-section for the process $e^+e^-\rightarrow f^+f^-$
is given by (at tree level)
\be
\sigma = {s\over 3\cdot 256\pi}\sum_{h_f,h_e}|F(h_f,h_e)|^2
\ee
with
\be
F(h_f,h_e)=-{4eq_f\over s}+
\sum_{a=Z,L_3,R_3}
{(v^f_a+h_f a^f_a)(v_a+h_e a_a)
\over {s-M_a^2+iM_a\Gamma_a}}
\ee
where $h_{f,e}=\pm 1$ are the helicities of $f$ and $e$ respectively, $q_f$
is the electric charge of $f$ ,
$v_{a}=v^e_{a}$,  $a_{a}=a^e_a$, with $a=Z,L_3,R_3$, and
$\Gamma_{a}$  are the widths of the neutral gauge bosons.
The partial widths of the $L_3$ and $R_3$ bosons corresponding to decays into
fermion-antifermion and $WW$ are given in sect. 12.

The forward-backward asymmetry in the present case is given by
\be
A_{FB}^{e^+e^-\to f^+ f^-}=\f 3 4
{{(1-P)\sum_{h_f,h_e}h_fh_e|F(h_f,h_e)|^2+2P\sum_{h_f}h_f|F(h_f,1)|^2}\over
{(1-P)\sum_{h_f,h_e}|F(h_f,h_e)|^2+2P\sum_{h_f}|F(h_f,1)|^2}}
\ee
where $P$ is the degree of longitudinal polarization of the
electron beam.
\par
The left-right asymmetry is given by
\be
A_{LR}^{e^+e^-\to f^+ f^-}=P{{\sum_{h_f,h_e}h_e|F(h_f,h_e)|^2}\over
{\sum_{h_f,h_e}|F(h_f,h_e)|^2}}
\ee
The notations are the same as for the forward-backward asymmetry.

In our study we consider also the $WW$ channel, with one $W$  decaying
leptonically and the other hadronically.  The reason for choosing this decay
channel is to get a clean signal to reconstruct the polarization of the $W$'s
(see for example \cite{fujii}). For the $e^+e^- \to WW$ channel the relevant
observables are:
\bea
&\dd{{d\sigma \over {d\cos\theta}}}(e^+ e^-\to W^+ W^-)\nn\\
& A_{LR}^{{ e^+ e^- \to W^+ W^-}}=(
\dd{{d\sigma \over {d\cos\theta}}}(P_{e}=+P)-
\dd{{d\sigma \over {d\cos\theta}}}(P_{e}=-P))/
\dd{{d\sigma \over {d\cos\theta}}}
\eea
where $\theta$ is the $e^+e^-$ center of mass scattering angle.
Assuming that the final $W$ polarization can be reconstructed by using the
$W$ decay distributions, it is convenient to consider the cross-sections for
$W_LW_L$, $W_TW_L$, $W_TW_T$ and the corresponding left-right asymmetries as
additional observables \cite{lays}.

In the $e^+e^-$ center of mass frame the angular distribution
$d\sigma/d\cos\theta$ and the left-right asymmetry read \cite{chife}
\bea
{{d\sigma}\over {d\cos\theta}}&=& {p\over 64\pi\sqrt{s}}
\Big{\{} a_W^4\left [{4\over M^2_W}+p^2\sin^2\theta
\left ( {1\over M_W^4}+{4\over t^2}\right )\right ]\\ \nn
&+&2F_1p^2\left [ {{4s}\over M^2_W}+
\left ( 3+{{sp^2}\over M^4_W}\right )\sin^2\theta\right ]\\ \nn
&+&F_1^{\prime}
\left [ 8\left ( 1+{{M^2_W}\over t}\right )+
16 {{p^2}\over M^2_W}+{{p^2}\over s}\sin^2\theta
\left ( {{s^2}\over M_W^4}-2{s\over M_W^2}-4{s\over t}\right )
\right ] \Big{\}}\\
\eea
and
\bea
A_{LR}(\cos\theta)&=&-P
{p\over 64\pi\sqrt{s}}
\Big{\{} a_W^4\left [{4\over M^2_W}+p^2\sin^2\theta
\left ( {1\over M_W^4}+{4\over t^2}\right )\right ]\\ \nn
&+&2F_2p^2\left [ {{4s}\over M^2_W}+
\left ( 3+{{sp^2}\over M^4_W}\right )\sin^2\theta\right ]\\ \nn
&+&F_1^{\prime}
\Big [ 8\left ( 1+{{M^2_W}\over t}\right )+
16 {{p^2}\over M^2_W}\\ \nn
&+&{{p^2}\over s}\sin^2\theta
\left ( {{s^2}\over M_W^4}-2{s\over M_W^2}-4{s\over t}\right )
\Big ] \Big{\}} {\Big /}{{d\sigma}\over {d\cos\theta}}\\
\label{alr}
\eea
where
\bea
&p&={1\over 2}\sqrt{s}(1-4M^2_W/s)^{1/2}\\ \nn
&t&=M_W^2-{1\over 2}s[1-\cos\theta (1-4M_W^2/s)^{1/2}]
\eea
The quantity $a_W$ is given in (9.2) and
\bea
F_1&=&\Big(
{{2e^2}\over s}\Big)^2+
\sum_{a=Z,L_3,R_3}\Big[
(v_a^2+a_a^2)g^2_{aWW}{1\over {(s-M_a^2)^2+M_a^2\Gamma_a^2}}\nn\\
&-&
4{e^2\over s}v_ag_{aWW} {{s-M_a^2}\over {(s-M_a^2)^2+M_a^2\Gamma_a^2}}\Big]
\nn \\
&+&\sum^{a\neq b}_{a,b=Z,L_3,R_3}
 (v_av_b+a_a a_b)g_{aWW}g_{bWW}\nn\\
& & {{(s-M_a^2)(s-M_b^2)+
M_a\Gamma_a M_b\Gamma_b}
\over {[(s-M_a^2)^2+M_a^2\Gamma_a^2][(s-M_b^2)^2+M_b^2\Gamma_b^2]}}
\eea
\be
F_1^{\prime}=a^2_W [{-2{e^2}\over s}+\sum_{a=Z,L_3,R_3}g_{aWW}(v_a+a_a)
{{s-M_a^2}\over {(s-M_a^2)^2+M_a^2\Gamma_a^2}}]
\ee
\bea
F_2&=&-{4e^2\over s}\sum_{a=Z,L_3,R_3}a_a g_{aWW}{{s-M_a^2}\over
{(s-M_a^2)^2+M_a^2
\Gamma_a^2}}\nn \\
&+&2\sum_{a=Z,L_3,R_3}a_av_ag_{aWW}^2{1\over
{(s-M_a^2)^2+M_a^2\Gamma_a^2}}\nn\\
&+&\sum^{a\neq b}_{a,b=Z,L_3,R_3}
(a_av_b+v_aa_b)g_{aWW}g_{bWW}\nn\\
&&{{(s-M_a^2)(s-M_b^2)+M_a\Gamma_a
 M_b\Gamma_b}
\over {[(s-M_a^2)^2+M_a^2\Gamma_a^2][(s-M_b^2)^2+M_b^2\Gamma_b^2]}}
\eea
where $g_{ZWW}$ $g_{L_3WW}$ and $g_{R_3WW}$ are given in eq. (10.2).

The cross-sections for $W_LW_L$, $W_TW_L$, and $W_TW_T$ are:
\bea
{{d\sigma_{LL}}\over {d\cos\theta}}&=&{p\over 64\pi\sqrt{s}}
\Big{\{} {{a_W^4}\over 16 M_W^4}{1\over t^2}
[ s^3 (1+\cos^2\theta)-4M_W^4 (3s+4 M_W^2)\nn\\
&-&4(s+2M_W^2)p\sqrt {s} s \cos\theta  ]\sin^2\theta\nn\\
&+&{F_1\over 8 M^4_W}\sin^2\theta
(s^3-12sM_W^4-16M_W^6)\nn\\
&+&F_1^{\prime}\sin^2\theta {1\over {2t}} [
ps\sqrt{s}\cos\theta{1\over {2M_W^4}}(s+2M_W^2)\nn\\
&-&{1\over {4 M_W^4}}(s^3-12sM_W^4-16M_W^6) ]
\Big{\}}
\label{ll}
\eea
\bea
{{d\sigma_{TL}}\over {d\cos\theta}}&=&{{p}\over 64\pi\sqrt{s}}
\Big{\{} {{a_W^4}}{1\over {2t^2 M_W^2}}
[ s^2 (1+\cos^4\theta)+4M_W^4 (1+\cos^2\theta)\nn\\
&-&4(4p^2+s\cos^2\theta )p\sqrt {s}  \cos\theta
+2s(s-6M_W^2)\cos^2\theta-4sM_W^2 ]\nn\\
&+&4F_1s{p^2\over M_W^2}(1+\cos^2\theta)\nn\\
&+&2F_1^{\prime}
{{p\sqrt{s}}\over {tM_W^2}}[
\cos\theta (4p^2+s\cos^2\theta)-
2p\sqrt{s}(1+\cos^2\theta)]
\Big{\}}
\label{tl}
\eea
\bea
{{d\sigma_{TT}}\over {d\cos\theta}}&=&{{p}\over {64\pi\sqrt{s}}}
\Big{\{} {{a_W^4}}{1\over {t^2 }}
[ s (1+\cos^2\theta)-2M_W^2-2p\sqrt {s}\cos\theta ]\sin^2\theta\nn\\
&+&4F_1{p^2}\sin^2\theta+F_1^{\prime}
{{\sin^2\theta}\over {2t}}[4p\sqrt{s}\cos\theta -8p^2]
\Big{\}}
\label{tt}
\eea
The left-right asymmetries for longitudinal and/or transverse polarized
$W$ can be easily obtained as in eq. (\ref{alr}) by substituting $F_1$ by
$F_2$
in eqs. (\ref{ll}), (\ref{tl}), (\ref{tt}), and dividing by the corresponding
differential cross-section.
\begin{figure}
\epsfysize=8truecm
\centerline{\epsffile{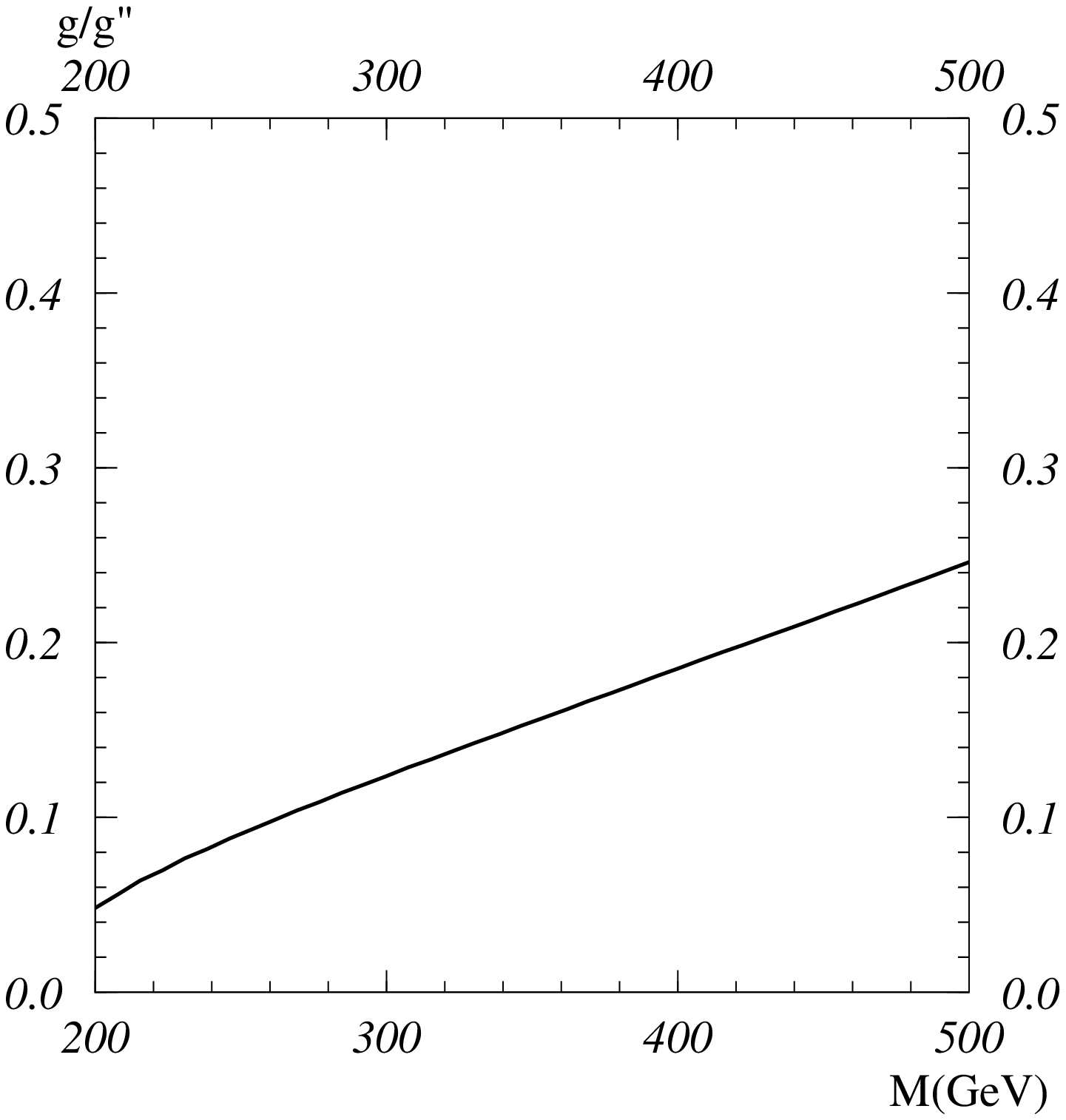}}
\noindent
{\bf Fig. 4} - {\it 90\% C.L. contour on the plane ($M$, $g/g''$) from LEP2.
The limits are obtained
considering $\sqrt{s}=175~ GeV$ and with an integrated luminosity of
$500 pb^{-1}$, combining the deviations of $M_W$, $\sigma^\mu$, $\sigma^h$,
$A_{FB}^{\mu}$, $A_{FB}^b$.}
\end{figure}
At LEP2 we can add to the previous observables the $W$ mass measurement,
coming from the $e^+e^- \to WW$ channel.
In Fig. 4 we show a 90\% C.L. contour plot in the parameter space of the
model.
The limits are obtained considering $\sqrt{s}=175~ GeV$ and an integrated
luminosity of $500 pb^{-1}$, combining the deviations of $M_W$, $\sigma^\mu$,
$\sigma^h$, $A_{FB}^{\mu}$, $A_{FB}^b$. For $M_W$ we assume a total error
(statistical and systematic) $\Delta M_W=50~MeV$. For $\sigma^h$ the total
error assumed is 2\%. For the other observable quantities we assume only
statistical errors. If the possibility of having polarized beams at LEP2
 is considered, the improvement with respect to the unpolarized case is only
marginal.
Also, considering the option of LEP2
at $\sqrt{s}=190~ GeV$ does not substantially alter the result.
The comparison with LEP bounds (see Fig. 1)
shows that LEP2 will not improve considerably the existing limits.
Of course one has to be careful in this comparison, since in the case
of LEP we have experimental values, whereas for LEP2 case the limits are
obtained by using deviations from the SM results.

To further test the model is necessary to consider higher energy colliders.
We study two options for a high energy $e^+e^-$ collider:
$\sqrt{s}=500~GeV$ with an integrated luminosity of $20 fb^{-1}$ and
$\sqrt{s}=1~TeV$ with an integrated luminosity of $80 fb^{-1}$.

\begin{figure}
\epsfysize=8truecm
\centerline{\epsffile{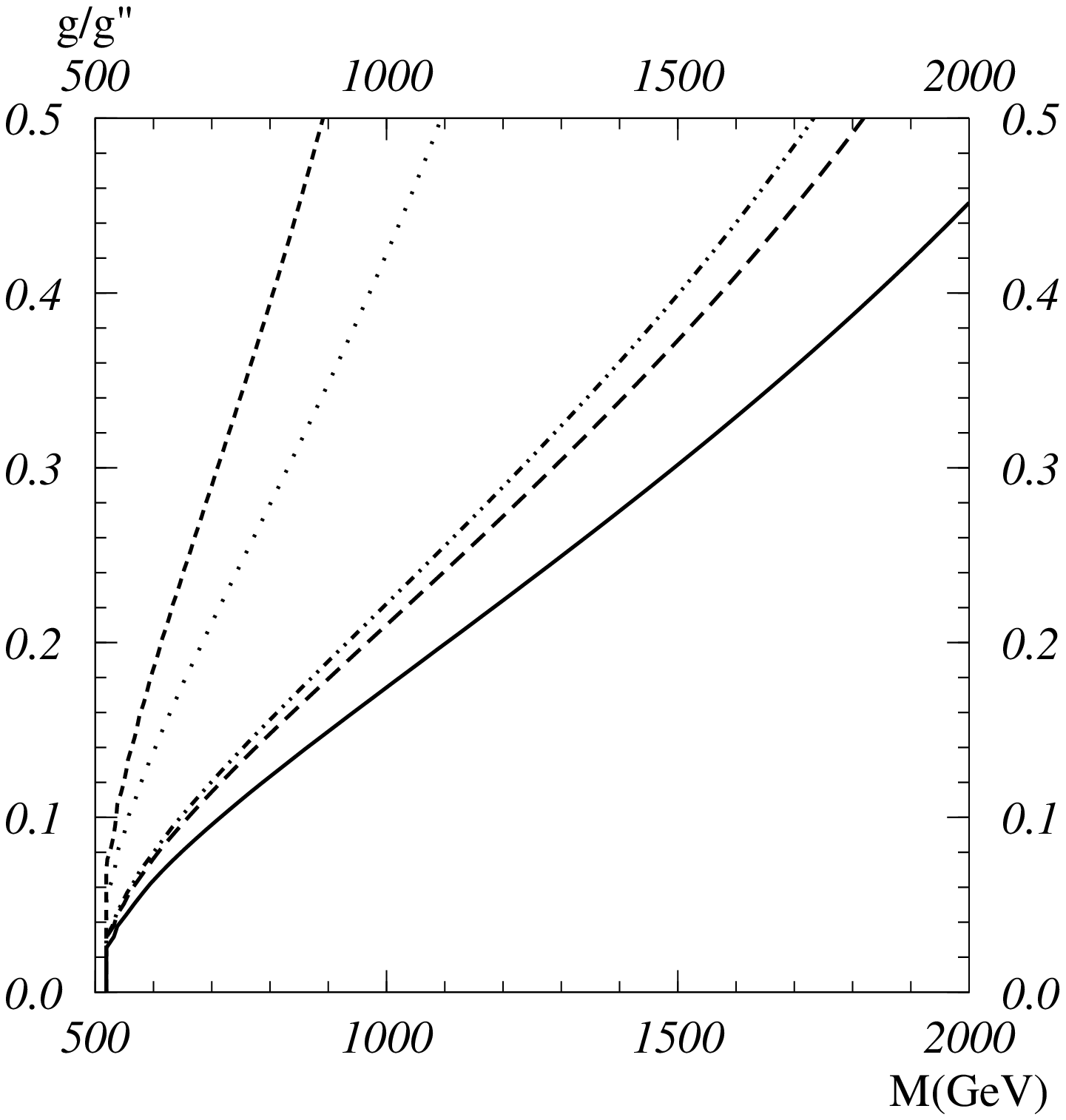}}
\noindent
{\bf Fig. 5} - {\it 90\% C.L. contour on the plane ($M$, $g/g''$) from
$e^+e^-$
at $\sqrt{s}=500~GeV$ with an integrated luminosity of $20 fb^{-1}$ for
various
observables. The dashed-dotted line
represents the limit from $\sigma^h$ with an assumed error of 2\%; the dashed
line near to the preceeding one is $\sigma^\mu$ (error 1.3\%); the dotted
line
is $A_{FB}^\mu$ (error 0.5\%); the uppermost dashed line is $A_{FB}^b$ (error
0.9\%). The continuous line represents the combined limits.}
\end{figure}

\begin{figure}
\epsfysize=8truecm
\centerline{\epsffile{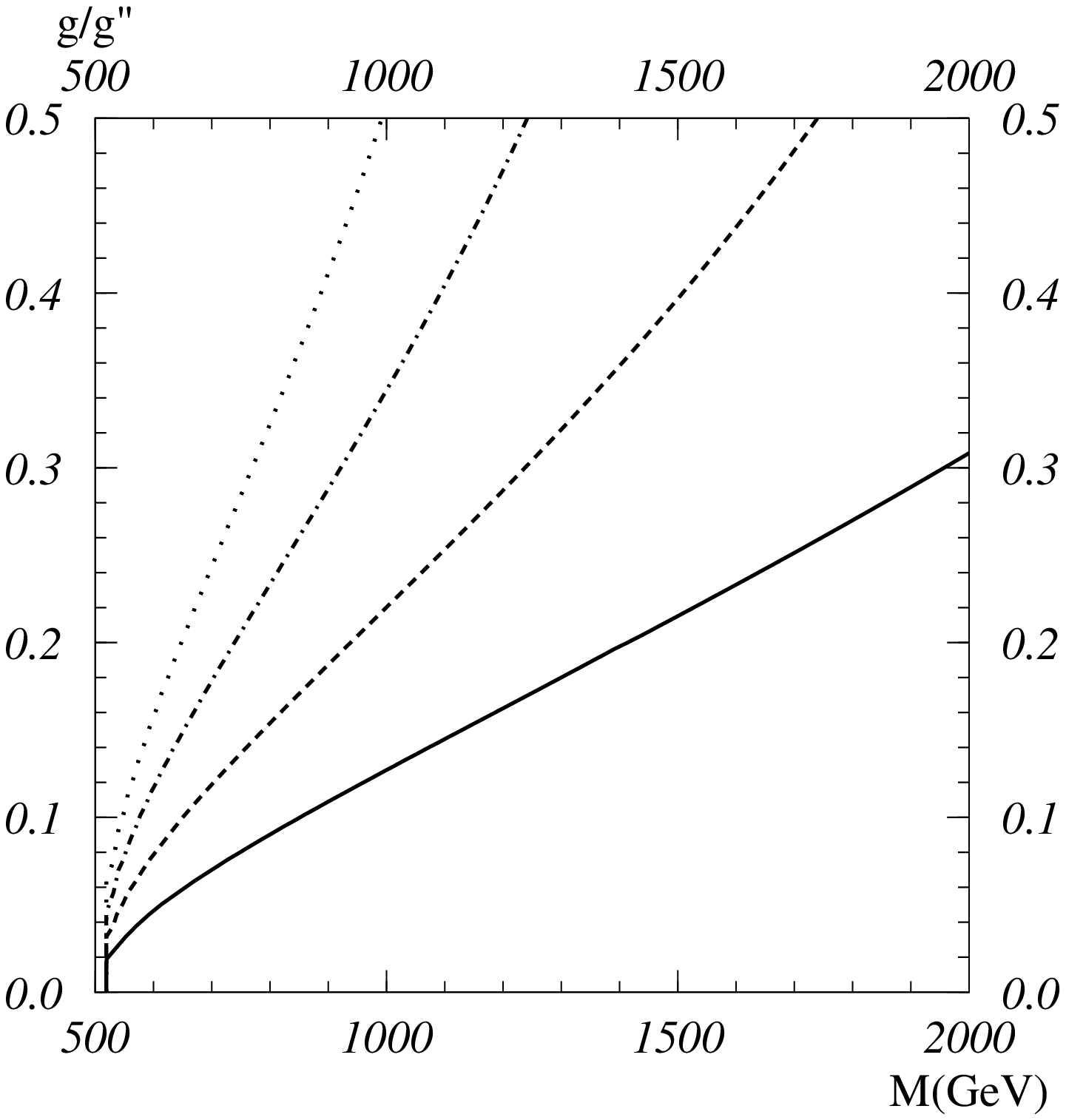}}
\noindent
{\bf Fig. 6} - {\it 90\% C.L. contour on the plane ($M$, $g/g''$) from
$e^+e^-$
at $\sqrt{s}=500~GeV$ with an integrated luminosity of $20 fb^{-1}$ and a
polarization of 0.5 for various observables. The dashed-dotted line
represents
the limit from $A_{LR}^\mu$ with an
assumed error of 0.6\%; the dashed line is $A_{LR}^h$ (error 0.4\%); the
dotted
line is $A_{LR}^b$ (error 1.1\%). The continuous is obtained by combining
the polarized and the unpolarized observables:
 $\sigma^h$, $\sigma^\mu$,
$A_{FB}^\mu$, $A_{FB}^b$, $A_{LR}^\mu$, $A_{LR}^h$, $A_{LR}^b$.}
\end{figure}

In Fig. 5 we present the 90\% C.L. contour on the plane
($M$, $g/g''$) from $e^+e^-$ at $\sqrt{s}=500~GeV$ with
an integrated luminosity of $20 fb^{-1}$ for various
observables. The dashed-dotted line
represents the limit from $\sigma^h$; the dashed
line near to the preceeding one is $\sigma^\mu$, the dotted line
is $A_{FB}^\mu$ and the uppermost dashed line is $A_{FB}^b$.
 As it is evident more stringent bounds come from the
cross-section measurements. Asymmetries give less restrictive bounds due to
a compensation between the $L_3$ and $R_3$ exchange.
By combining all the deviations in the previously considered
observables we get the limit shown by the continuous line.

Polarized electron beams
allow to get further limit in the parameter space as shown in Fig. 6.
We neglect the error on the measurement of the polarization and use a
polarization value equal to 0.5.
The dashed-dotted line represents the limit from $A_{LR}^\mu$,
the dashed line from $A_{LR}^h$, the dotted line from $A_{LR}^b$.
Combining all the polarized and unpolarized beam observables we get the
bound shown by the continuous line. In conclusion
we get a substantial improvement with respect to the LEP bounds,
even without polarized beams.

\begin{figure}
\epsfysize=8truecm
\centerline{\epsffile{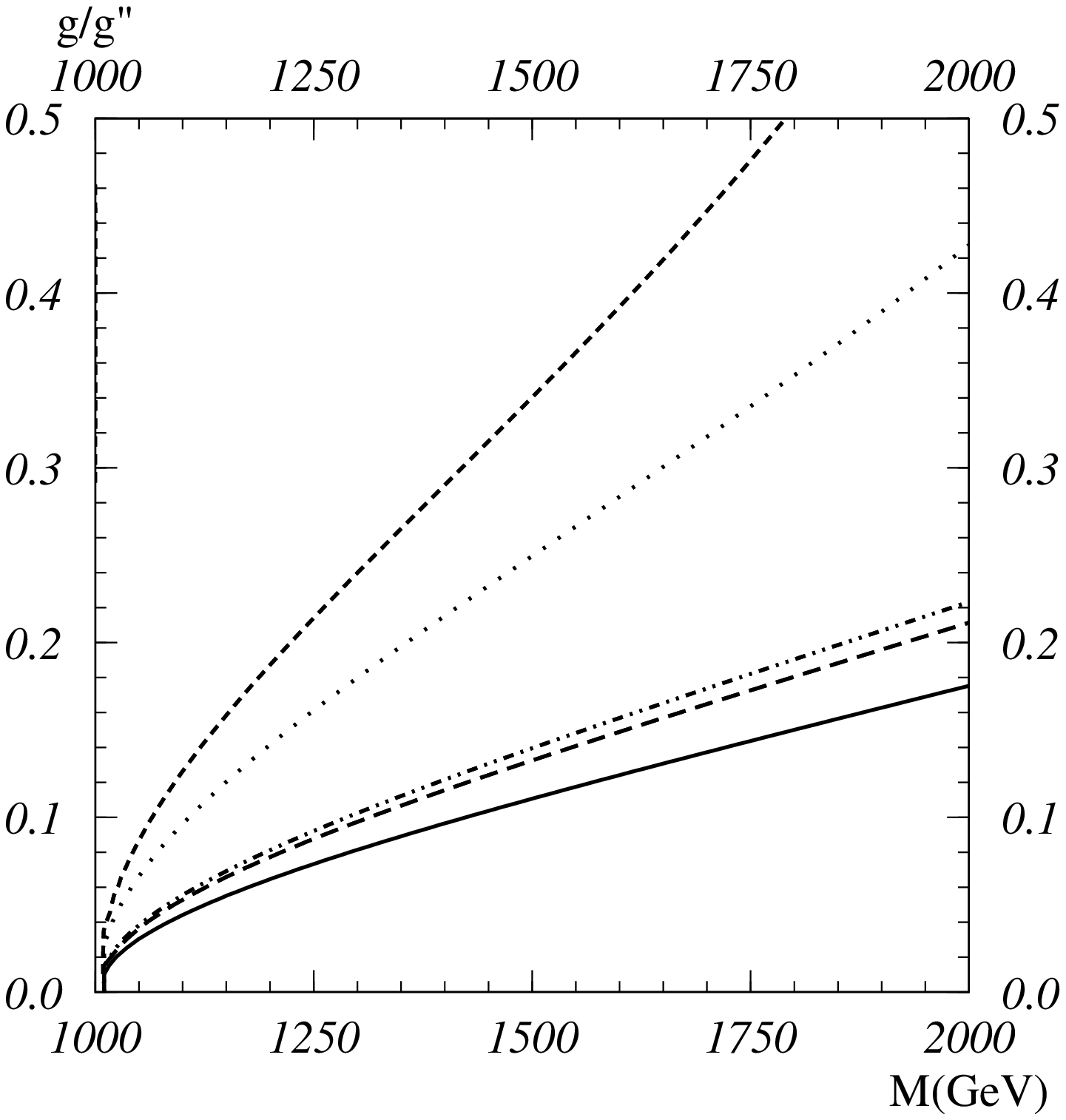}}
\noindent
{\bf Fig. 7} - {\it 90\% C.L. contour on the plane ($M$, $g/g''$) from
$e^+e^-$
at $\sqrt{s}=1000~GeV$ with an integrated luminosity of $80 fb^{-1}$ for
various
observables. The dashed-dotted line
represents the limit from $\sigma^h$ with an assumed error of 2\%; the dashed
line near to the preceeding one is $\sigma^\mu$ (error 1.3\%); the dotted
line
is $A_{FB}^\mu$ (error 0.5\%); the uppermost dashed line is $A_{FB}^b$ (error
0.9\%). The continuous line represents the combined limits.}
\end{figure}

\begin{figure}
\epsfysize=8truecm
\centerline{\epsffile{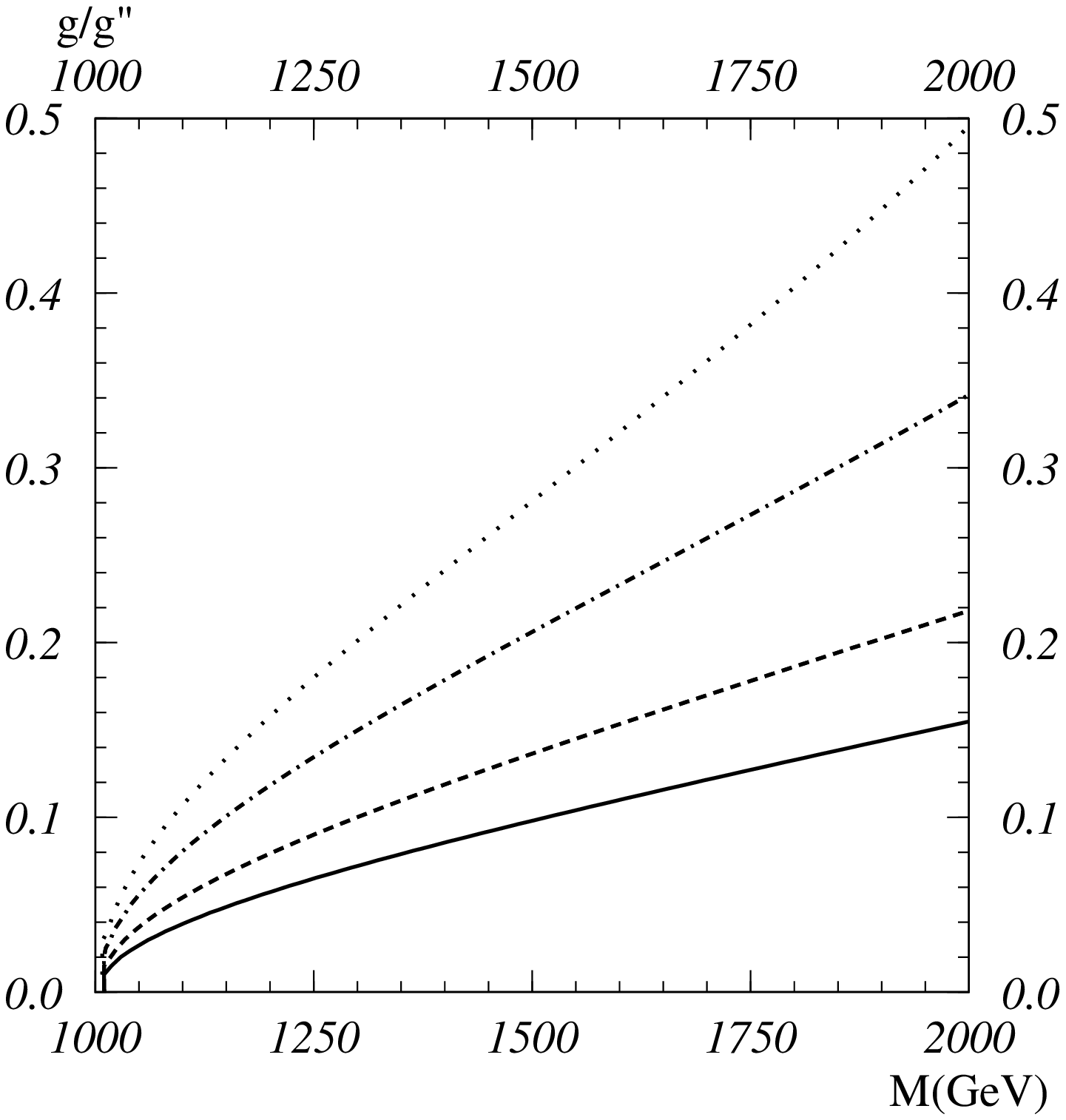}}
\noindent
{\bf Fig. 8} - {\it 90\% C.L. contour on the plane ($M$, $g/g''$) from
$e^+e^-$
at $\sqrt{s}=1000~GeV$ with an integrated luminosity of $80 fb^{-1}$ and a
polarization of 0.5 for various observables. The dashed-dotted line
represents
the limit from $A_{LR}^\mu$ with an
assumed error of 0.6\%; the dashed line is $A_{LR}^h$ (error 0.4\%); the
dotted
line is $A_{LR}^b$ (error 1.1\%). The continuous line is obtained by
combining
the polarized and the unpolarized observables:
 $\sigma^h$, $\sigma^\mu$,
$A_{FB}^\mu$, $A_{FB}^b$, $A_{LR}^\mu$, $A_{LR}^h$, $A_{LR}^b$.}
\end{figure}

The previous analysis has been repeated at
$\sqrt{s}=1~TeV$ with an integrated luminosity of $80 fb^{-1}$.
The results are shown in Figs. 7 and 8.

\begin{figure}
\epsfysize=8truecm
\centerline{\epsffile{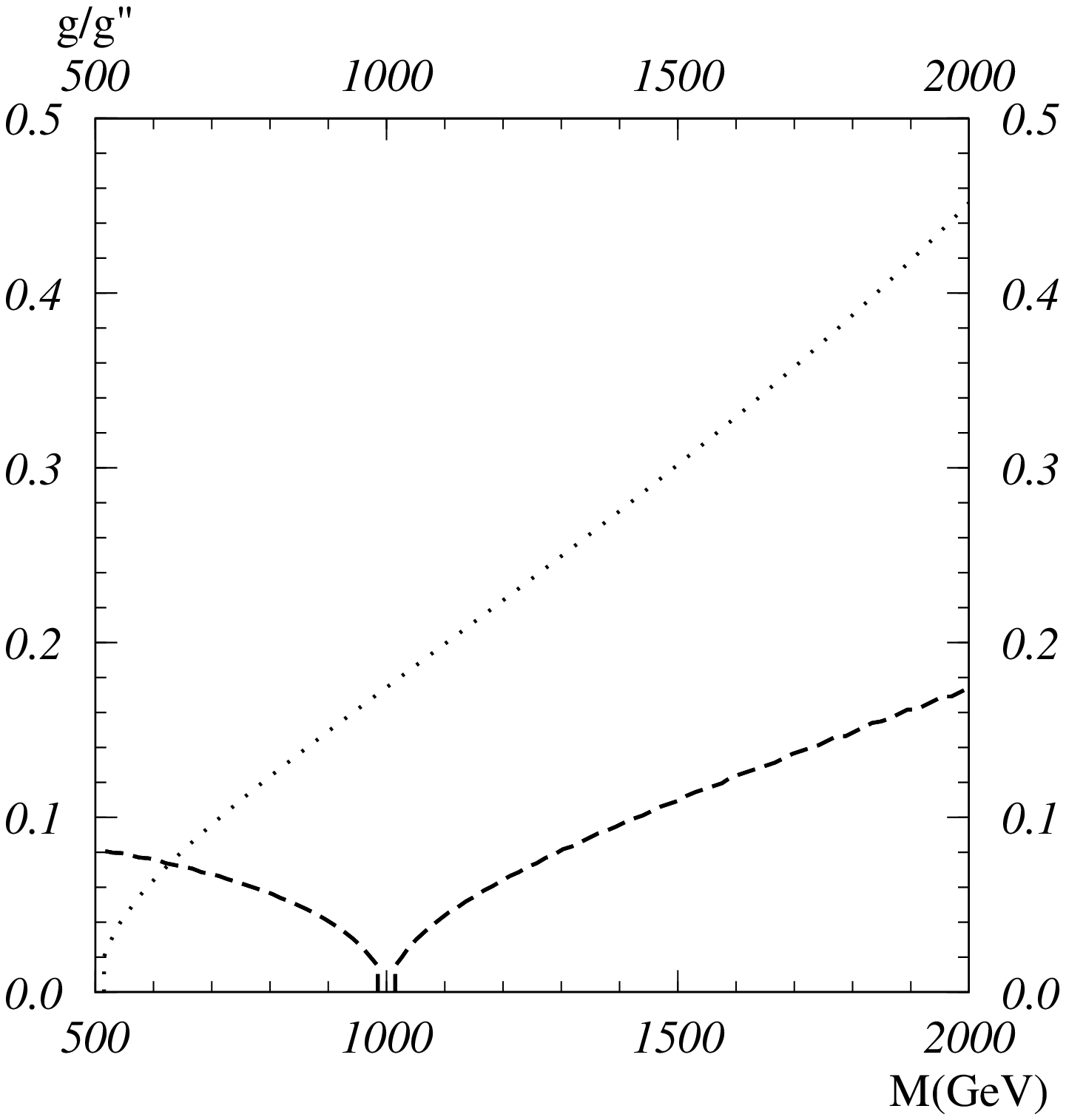}}
\noindent
{\bf Fig. 9} - {\it 90\% C.L. contour on the plane ($M$, $g/g''$) from
$e^+e^-$ at two $\sqrt{s}$ values:  the dotted line represents
the limit from the
combined unpolarized observables at $\sqrt{s}=500~GeV$ with an
integrated luminosity of $20 fb^{-1}$; the dashed line is the limit
from the combined unpolarized observables at $\sqrt{s}=1000~GeV$
with an integrated luminosity of $80 fb^{-1}$.}
\end{figure}

\begin{figure}
\epsfysize=8truecm
\centerline{\epsffile{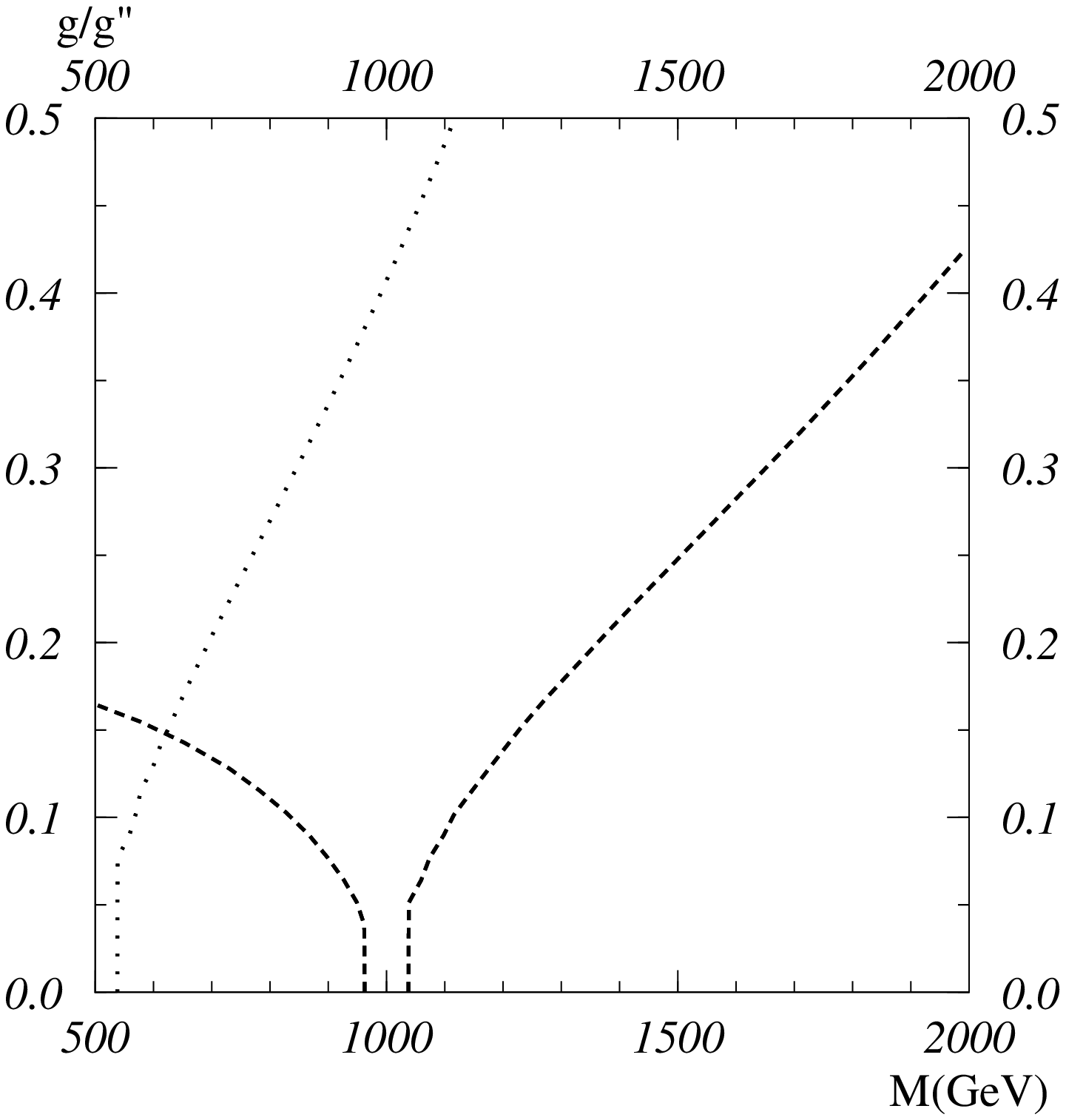}}
\noindent
{\bf Fig. 10} - {\it 90\% C.L. contour on the plane ($M$, $g/g''$) from
$WW$ differential cross-section and the corresponding left-right asymmetries,
considering also the $W$ polarization reconstruction. The dotted line
represents
the limit from $\sqrt{s}=500~GeV$ with an integrated luminosity of $20
fb^{-1}$;
the dashed line is the limit from $\sqrt{s}=1000~GeV$ with an integrated
luminosity of $80 fb^{-1}$.}
\end{figure}

In Fig. 9 we show a combined picture of the
 90\% C.L. contours on the plane ($M$, $g/g''$) from $e^+e^-$
at two values of $\sqrt{s}$.  The dotted
line represents the limit from the combined unpolarized observables at
$\sqrt{s}=500~GeV$ with an integrated luminosity of $20 fb^{-1}$; the
dashed line is the limit from the combined unpolarized observables
at $\sqrt{s}=1000~GeV$ with an integrated luminosity of $80 fb^{-1}$.
As expected increasing the energy of the collider and rescaling the
integrated luminosity result in stronger bounds on the
parameter space.

We have then studied the $WW$ final state, considering the observables
given in eq. (14.6). In Fig. 10 we show the plot from the combined $WW$
observables. An angular cut has been imposed on $W$ scattering angle ($| \cos
\theta | \leq 0.95$) and 18 angular bins have been considered.
We have assumed an overall detection efficiency of $10\%$ including
the branching ratio $B=0.29$ and a loss of luminosity from beamstrahlung.
All
these new bounds do not alter the strong limits
obtained using the fermion final state. This is  because, as we
have already noticed, the degenerate model has no strong
enhancement of the $WW$ channel, present in the usual
strong electroweak models.

\resection{Degenerate BESS at hadron colliders}

The $e^+ e^-$ colliders give the possibility to explore the neutral sector of
symmetry breaking by the production of the neutral vector and axial vector
gauge bosons of the model. Hadron colliders are complementary to $e^+ e^-$
machines, in the sense that they also allow to study the new charged vector
and
axial vector resonances.

The physics of large hadron colliders has been extensively discussed in a
number of papers (see for example \cite{LHC} and references therein); such a
machine will be able either to discover the new resonances or to constrain
the physical region left unconstrained by previous data.

Let us consider first the case in which no new resonances are discovered.
In this case limits can be imposed on the parameter space of the model. As a
preliminary analysis we can consider the total cross-section of
$pp \to L^\pm,
W^\pm \to
\ell \nu$ ($\ell=e,\mu$), which has a clear signature and a large number of
events, to be compared with the standard model production of $\ell \nu$.
We have calculated the total cross-section $pp \to L^\pm, W^\pm \to
\ell \nu$ and, by comparing with the SM background, we have
 obtained a contour plot at 90\% C.L. in the two variables
$M$ and $g/g''$, shown in Fig. 11.
\begin{figure}
\epsfysize=8truecm
\centerline{\epsffile{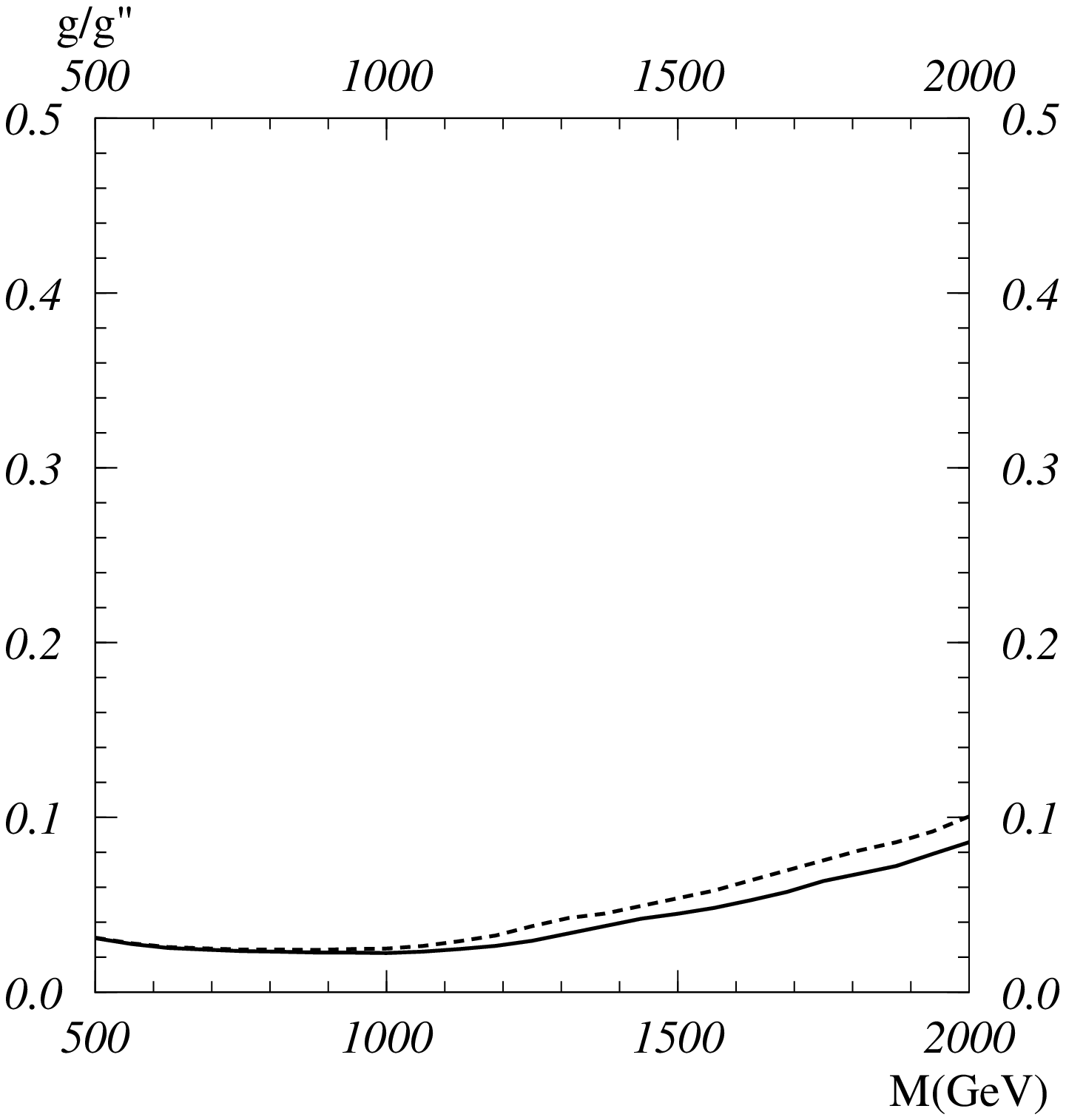}}
\noindent
{\bf Fig. 11} - {\it 90\% C.L. contour on the plane ($M$, $g/g''$) from
total cross-section of $pp \to L^\pm, W^\pm \to
\ell \nu$ ($\ell=e,\mu$). We have assumed a systematical error of
5\% in the cross-section and the statistical error obtained considering a
luminosity of $10^{34} cm^{-2} s^{-1}$ (continuous line) or a luminosity of
$10^{33} cm^{-2} s^{-1}$ (dashed line) and one year run at LHC ($10^7$ s).}
\end{figure}

The applied cuts were $\vert p_{t\mu}\vert > min(M_{L^\pm}/2 -50~ GeV, 400
{}~ GeV)$ in order to maximize the deviation of BESS model cross-section with
respect to the Standard Model one. In this analysis we do not
optimize cuts and an improvement is still possible studying in more detail
specific cases. We have assumed a systematical error of
5\% in the cross-section and the statistical error obtained  considering a
luminosity of $10^{34} cm^{-2} s^{-1}$ (continuous line) or a luminosity of
$10^{33} cm^{-2} s^{-1}$ (dashed line) and one year run ($10^7$ s) at LHC
($\sqrt{s}=14~TeV$).
The new resonances of the model can be discovered directly for a wide range
of values of the parameter space of the model. The discovery limit in the
mass of the resonance depends on the value of $g/g''$.
For example if $g/g''=0.1$, the
resonance is visible over the background at least up to 2 $TeV$, in the
channel $pp \to \mu \nu$.

In Figs. 12-14 we show the differential distribution of events at LHC of
$pp \to L^\pm,
W^\pm \to
\mu \nu$  in the transverse momentum of the muon for
different values of
$M_{L^\pm}$ and $g/g''$. As stated before we choose this channel due to the
clean signature and the large cross-section. The events where simulated using
Pythia Montecarlo \cite{phy}. A rough simulation of the detector was
also performed. The energy of the leptons was smeared according to
\be
\frac{\Delta E}{E} = 15\%
\ee
and the error in the 3-momentum determination was assumed of 5\%.

\begin{figure}
\epsfysize=8truecm
\centerline{\epsffile{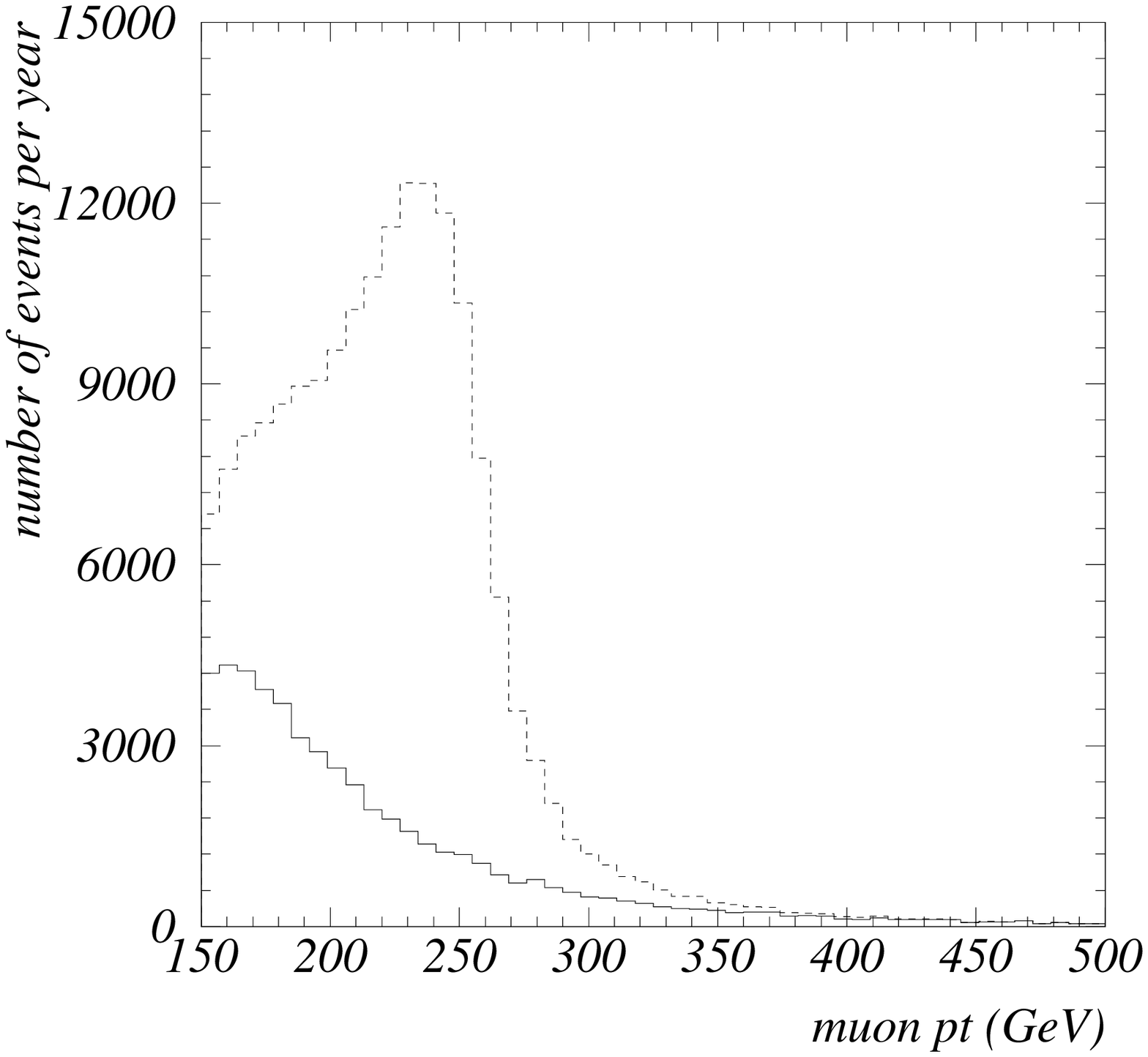}}
\smallskip
\noindent
{\bf Fig. 12} - {\it Differential distribution of $pp \to L^\pm, W^\pm \to
\mu \nu$ events at LHC with a luminosity of
$10^{34} cm^{-2} s^{-1}$, for $M_{L^{\pm}}=500~GeV$, $g/g''=0.15$.
The following cuts have been applied: $\vert p_{t\mu}\vert ,
\vert p_{t~miss}\vert >150~GeV$. The continuous line represents the
Standard Model background while the dashed one is the degenerate BESS model
expectation.}
\end{figure}

In particular in Fig. 12 a spectacular case is presented for a low resonance
mass $M_{L^{\pm}}=500~GeV$ and $g/g''=0.15$. The total $L^{\pm}$ width is
$\Gamma_{L^{\pm}}=0.907~GeV$, with the corresponding $B(L^+\to\mu\nu)=8.5\times
10^{-2}$. The following cuts have been applied: $\vert p_{t\mu}\vert ,
\vert p_{t~miss}\vert >150~GeV$. The number of signal events per year
is approximately 128000, the corresponding background consists of 51500
events.
\begin{figure}
\epsfysize=8truecm
\centerline{\epsffile{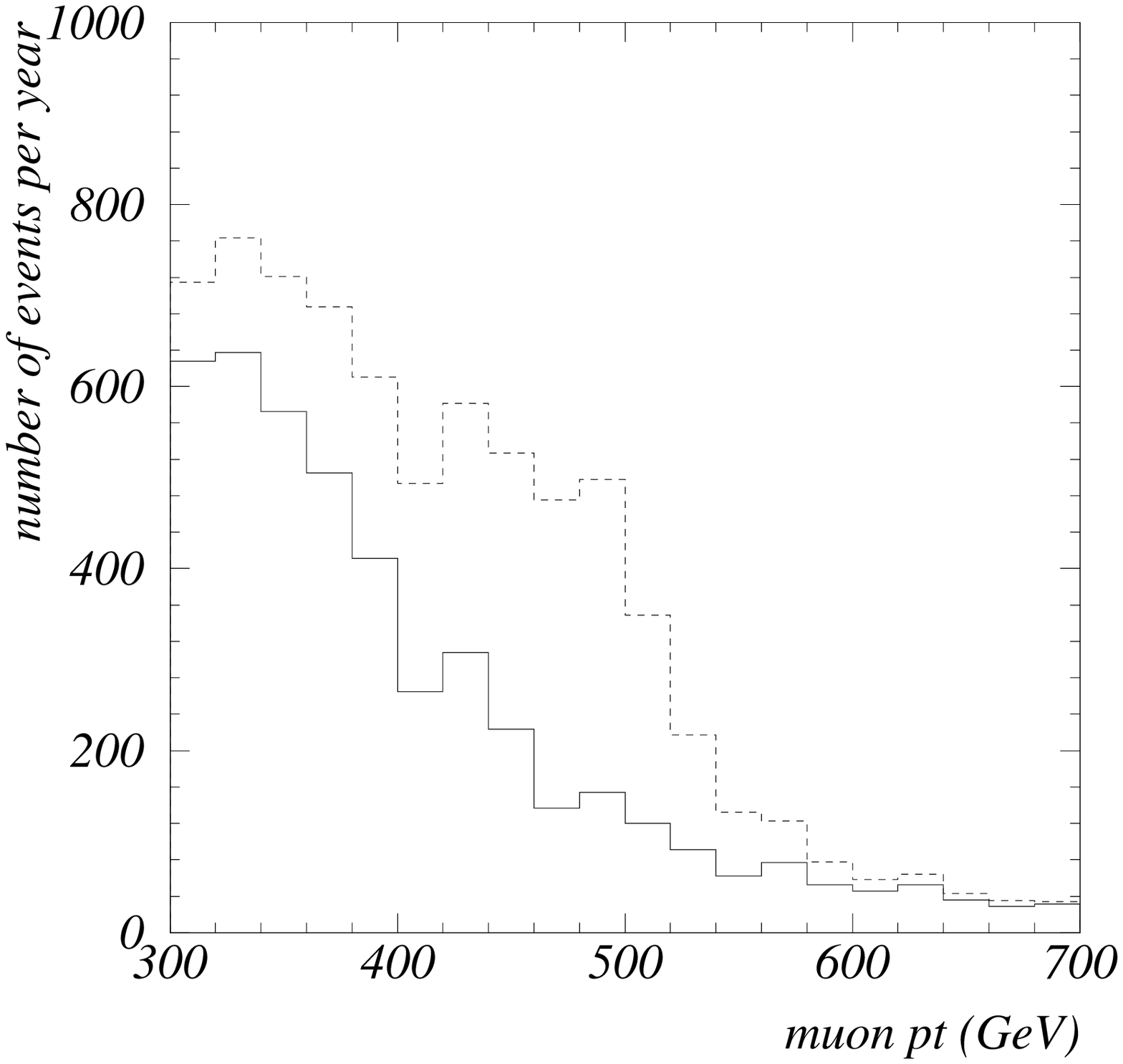}}
\noindent
{\bf Fig. 13} - {\it Differential distribution of $pp \to L^\pm, W^\pm \to
\mu \nu$ events at LHC with a luminosity of
$10^{34} cm^{-2} s^{-1}$, for $M_{L^{\pm}}=1~TeV$, $g/g''=0.075$.
The following cuts have been applied: $\vert p_{t\mu}\vert ,
\vert p_{t~miss}\vert >300~GeV$. The continuous line is the
Standard Model background, the dashed line represents the degenerate BESS
model
signal.}
\end{figure}

\begin{figure}
\epsfysize=8truecm
\centerline{\epsffile{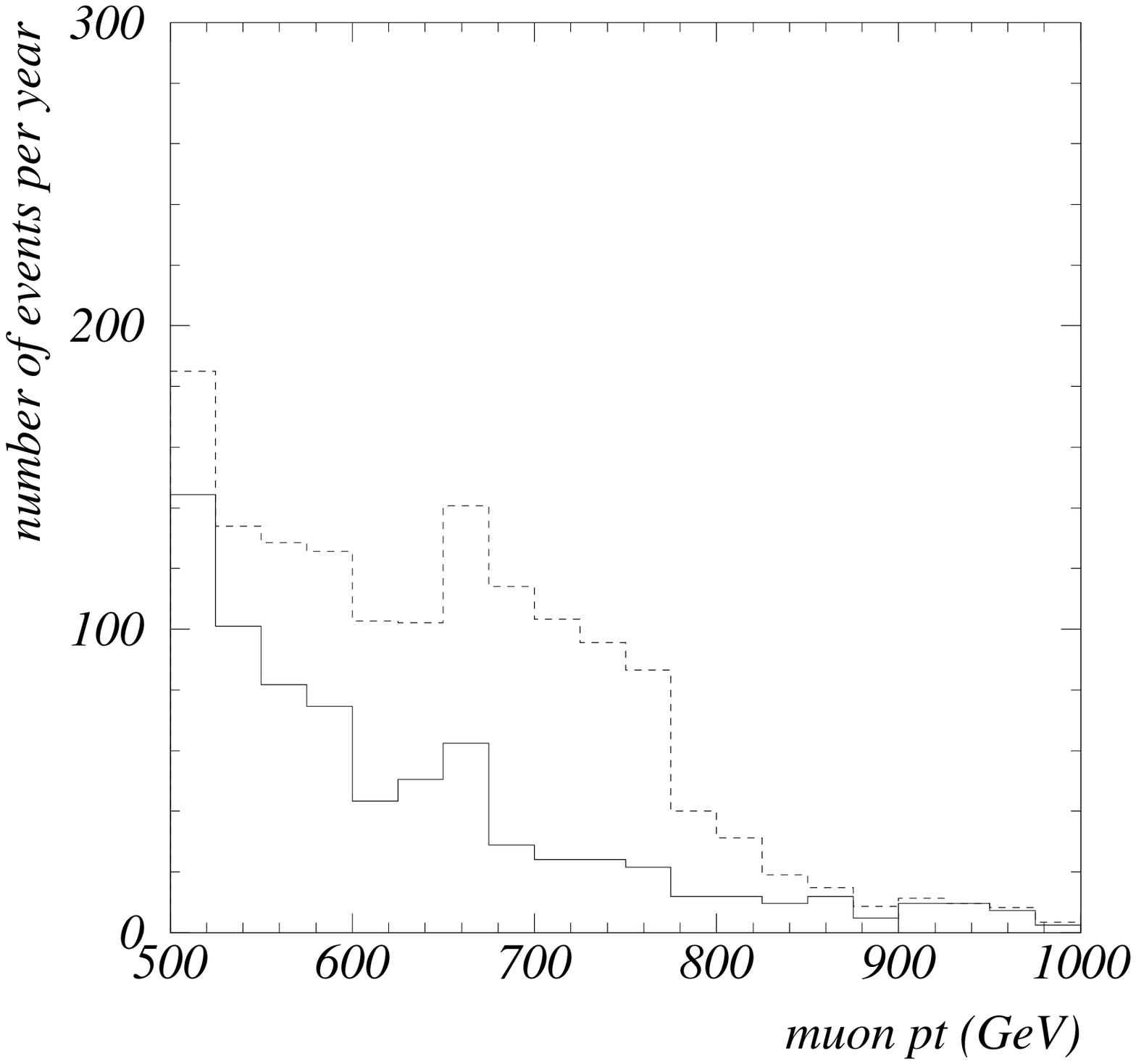}}
\noindent
{\bf Fig. 14} - {\it Differential distribution of $pp \to L^\pm, W^\pm \to
\mu \nu$ events at LHC with a luminosity of
$10^{34} cm^{-2} s^{-1}$, for $M_{L^{\pm}}=1.5~TeV$, $g/g''=0.1$.
The following cuts have been applied: $\vert p_{t\mu}\vert >400~GeV$,
$\vert p_{t~miss}\vert >400~GeV$. The continuous line represents the
Standard Model background while the dashed line is the degenerate
BESS model expectation.}
\end{figure}

In Fig. 13 we show a case corresponding to $M_{L^{\pm}}=1~TeV$, $g/g''=0.075$
and $\Gamma_{L^{\pm}}=0.454~GeV$. The following cuts have been applied:
$\vert
p_{t\mu}\vert , \vert p_{t~miss}\vert >300~GeV$ and $E_{miss}>100~GeV$.
The number of signal events per year is approximately 2800, the corresponding
background consists of 4600 events.

In Fig. 14 we show a case corresponding to $M_{L^{\pm}}=1.5~TeV$, $g/g''=0.1$
and $\Gamma_{L^{\pm}}=1~GeV$. The following cuts have been applied: $\vert
p_{t\mu}\vert , \vert p_{t~miss}\vert >400~GeV$ and $E_{miss}>200~GeV$.
The number of signal events per year is approximately 850, the corresponding
background consists of 1500 events. The statistical significance of the
signal
is $S/\sqrt B=22$.

Notice that, the reconstruction of resonance mass, requires a careful study
of
the experimental setup, due to the smallness of the resonance width.

In this preliminary study we did not consider the production and decay
of the corresponding neutral resonances of the model.

\resection{Conclusions}

We have discussed an effective theory describing new vector and axial vector
resonances within the scheme of a strong electroweak breaking sector.
We have shown
that the model has a symmetry which is larger than the one requested by
construction. No Higgs particles are required in this effective theory, and
moreover the enlarged symmetry guarantees that even with a relatively low
energy strong electroweak resonant sector, the severe constraints coming from
experimental data, in particular from LEP, are respected.

The new vector and axial vector particles are degenerate in mass (at the
leading order) and their virtual effects are suppressed. In the low energy
limit ($M\to \infty$ with the gauge coupling of the new resonances fixed) the
new particles are decoupled due to the extended symmetry $[SU(2)\otimes
SU(2)]^3$ and we classically obtain the effective lagrangian of the standard
model. If we parameterize the deviations from the SM at low energy in terms
of the $\epsilon$ parameters, we obtain a deviation from the standard model
values only in the next-to-leading order, with a double suppression factor
$M_Z^2/M^2$ and $(g/g'')^2$.

A well known feature of the usual strong interacting models is the
relevance of the $WW$ final state. Our model is in this respect different, as
the $WW$ final state is on the same footing with the fermionic final state.
The
reason is that the longitudinal parts of the $W$'s, related via the
equivalence
theorem to the absorbed Goldstone bosons, are decoupled from the new
resonances.
This is  due to the absence of coupling between
$U$ and $L$, $R$ (see sect. 2).

For what concerns virtual effects of the new resonances, the model has almost
no
deviations at low energy
with respect to the SM. The situation is completely different if the
direct production of the new resonances is considered. This will be possible
with the next generation of colliders.

In the case of  $e^+e^-$ colliders, LEP2 will not improve the
existing bounds from LEP and Tevatron.
A substantial improvement can be obtained from
higher energy electron-positron colliders, even without considering
polarized beams. The most stringent bounds come from cross-section
measurements,
while asymmetries are less restrictive due to compensations between the two
neutral resonances.

High energy hadron colliders, as LHC, will allow to study also the new
charged
resonances. If the new particles are not discovered, stringent bounds on the
model can be imposed by the study of cross-sections as $pp \to \ell \nu$,
in a way similar to the one in which bounds  are
searched for at Tevatron. The direct observation of the new resonances is
possible in a wide window of the parameter space of the model, up to the
$TeV$
range, in some case with a spectacular number of events over the background.

\end{document}